\documentclass[11pt,a4paper]{emulateapj}
 \usepackage{url}
\usepackage{amsmath}
\usepackage{xcolor}
\usepackage{multirow}
\usepackage{graphicx}
\usepackage[export]{adjustbox}
\usepackage{psfig}
\usepackage[flushleft]{threeparttable}

\usepackage[symbol*]{footmisc}
\DefineFNsymbolsTM{myfnsymbols}{
 \textasteriskcentered *
  \textdagger    \dagger
  \textsection   \mathsection
  \textbardbl    \|%
  \textparagraph \mathparagraph 
}%
\setfnsymbol{myfnsymbols}
 \begin{document}

\title{extended X-ray study of M49: the frontier of the Virgo Cluster}
\author{Y. Su\altaffilmark{$\ddagger$1,2}}
\author{R. P. Kraft\altaffilmark{2}}
\author{P. E. J. Nulsen\altaffilmark{2,3}}
\author{C. Jones\altaffilmark{2}}
\author{T. J. Maccarone\altaffilmark{4}}
\author{F. Mernier\altaffilmark{5,6,7}}
\author{L. Lovisari\altaffilmark{2}}
\author{A. Sheardown\altaffilmark{8}}
\author{S. W. Randall\altaffilmark{2}}
\author{E. Roediger\altaffilmark{8}}
\author{T. M. Fish\altaffilmark{8}}
\author{W. R. Forman\altaffilmark{2}}
\author{E. Churazov\altaffilmark{9}}

\affil{$^1$Department of Physics and Astronomy, University of Kentucky, 505 Rose Street, Lexington, KY, 40506, USA}
\affil{$^2$Harvard-Smithsonian Center for Astrophysics, 60 Garden Street, Cambridge, MA 02138, USA}
\affil{$^3$ICRAR, University of Western Australia, 35 Stirling Highway, Crawley,
WA 6009, Australia}
\affil{$^4$Department of Physics, Texas Tech University, Box 41051, Lubbock, TX 79409-1051, USA}
\affil{$^5$MTA-E\"{o}tv\"{o}s University Lendulet Hot Universe Research Group, P\'{a}zm\'{a}ny P\'{e}ter s\'{e}t\'{a}ny 1/A, Budapest, 1117, Hungary}
\affil{$^6$
Institute of Physics, E\"{o}tv\"{o}s University, P\'{a}zm\'{a}ny P\'{e}ter s\'{e}t\'{a}ny 1/A, Budapest, 1117, Hungary}
\affil{$^7$SRON Netherlands Institute for Space Research, Sorbonnelaan 2, 3584 CA Utrecht, The Netherlands}
\affil{$^8$E.A. Milne Centre for Astrophysics, Department of Physics and Mathematics, University of
Hull, Hull, HU6 7RX, United Kingdom}
\affil{$^9$Max Planck Institute for Astrophysics, Karl-Schwarzschild-Str. 1, 85741, Garching, Germany}
\altaffiltext{$\ddagger$}{Email: ysu262@uky.edu}


\begin{abstract}
The M49 group, resident outside the virial radius of the Virgo cluster, is falling onto the cluster from the south. We report results from deep {\sl XMM-Newton} mosaic observations of M49. Its hot gas temperature is 0.8\,keV at the group center and rises to 1.5\,keV beyond the brightest group galaxy (BGG). The group gas extends to radii of $\sim300$\,kpc to the north and south. The observations reveal a cold front $\sim20$\,kpc north of the BGG center and an X-ray bright stripped tail 70\,kpc long and 10\,kpc wide to the southwest of the BGG. We argue that the atmosphere of the infalling group was slowed by its encounter with the Virgo cluster gas, causing the BGG to move forward subsonically relative to the group gas. We measure declining temperature and metallicity gradients along the stripped tail. The tail gas can be traced back to the cooler and enriched gas uplifted from the BGG center by buoyant bubbles, implying that AGN outbursts may have intensified the stripping process. We extrapolate to a virial radius of 740\,kpc and derive a virial mass of $4.6\times10^{13}\,M_\odot$ for the M49 group. Its group atmosphere appears truncated and deficient when compared with isolated galaxy groups of similar temperatures. If M49 is on its first infall to Virgo, the infall region of a cluster could have profound impacts on galaxies and groups that are being accreted onto galaxy clusters. Alternatively, M49 may have already passed through Virgo once.



\end{abstract}

\section{\bf Introduction}

Hierarchical structure formation is the cornerstone of modern cosmology.
As the most massive virialized systems in the Universe, clusters of galaxies have formed late with the vast bulk of their dark matter, hot gas, and member galaxies assembled since $z\sim0.5$ (e.g., Boylan-Kolchin et al.\ 2009). 
Galaxy groups are the building blocks of galaxy clusters, contributing 70\% of their mass (Berrier et al.\ 2009).  
Groups are continually being accreted
onto clusters through a web-like network of filamentary structures.
The radius that encloses a density contrast of 200$\times$ the critical density of the Universe, $R_{200}$,
which is approximately the virial radius, has often been used as the boundary of galaxy clusters. 
Recently, the splashback radius, defined as 
the first apocenter radius of galaxy orbits following their first pericenter passage,
has been proposed as a more physical extent of the cluster halo. 
Cosmological simulations suggest that the splashback radius is approximately $1.5\times$ the virial radius (More et al.\ 2015). 
We refer to the volume outside the virial radius, but within the splashback radius, as the cluster's infall region. 
The infall region provides unique distinguishing power for models of the hierarchical assembly of dark matter halos and the growth of galaxy clusters. Due to its low surface brightness, the infall region has rarely been explored in the X-ray.  

The intracluster medium (ICM) is a hot and ionized plasma radiating in X-rays via bremsstrahlung and constituting 90\% of the baryonic mass in galaxy clusters. The ICM can be stirred by infalling galaxies and groups, giving rise to a wealth of astrophysical processes.
Surface brightness edges in the ICM may form in the upstream region of infalling subclusters, separating gas of different entropy at so called ``cold fronts" (e.g., Abell 3667--Vikhilinin et al.\ 2001; NGC~1404--Su et al.\ 2017a,b). 
Bow shocks are expected in supersonic infall, creating a shock heated ICM. A highly turbulent ICM is expected to form in the wake behind the infalling object (Kraft et al.\ 2017; Roediger et al.\ 2015).  
High metallicity gas can be removed from infalling subclusters due to the dynamical pressure of the ICM, enriching the ICM over a large span of radii (Su et al.\ 2014). Conversely, the ICM has long been known to play a critical role in the evolution of member galaxies (e.g., Oemler 1974; Butcher \& Oemler 1978; Dressler 1980). In contrast to their counterparts in the field, cluster galaxies have quenched star formation. They also have less molecular gas and more disrupted morphologies. 
Additionally, since most galaxies reside in groups, the intragroup medium of infalling groups may affect the galaxy properties before the galaxy interacts with the rich cluster environment directly (often called pre-processing; Zabludoff et al.\ 1996; Fujita 2004; Bianconi et al.\ 2018). 

In the $\Lambda \rm CDM$ paradigm, the late assembly of clusters implies that there is at least one infalling galaxy group residing in the outskirts of each galaxy cluster (Haines et al.\ 2018). Dedicated X-ray observations have revealed a number of infalling groups such as the southern group in Hydra A (De Grandi et al.\ 2016), the northeastern group in Abell~2142 (Eckert et al.\ 2014, 2017), NGC~4389 in Coma (Neumann et al.\ 2003), and the southern group in Abell~85 (Ichinohe et al.\ 2015). These studies have greatly enhanced our knowledge of the astrophysical processes in the dynamically active cluster outskirts, i.e., shocks, bulk motion, turbulence, ram pressure stripping, and cluster plasma physics. 
However, none of the above examples are in the infall region of galaxy clusters. In this paper, we present our study of M49 (NGC~4472), a galaxy group residing beyond the virial radius of a galaxy cluster, Virgo, the true frontier of cluster evolution.



Virgo is the nearest galaxy cluster at a distance of $\sim16$\,Mpc.
Its ICM has been studied extensively using X-ray imaging spectroscopy ({\sl ROSAT}--B$\ddot{\rm o}$hringer et al.\ 1994; {\sl ASCA}--Shibata et al.\ 2001; {\sl XMM-Newton}--Urban et al.\ 2011; {\sl Suzaku}--Simionescu et al.\ 2015, 2017). 
The cluster has an average temperature of $kT=2.3$\,keV. Using {\sl Suzaku} observations out to the virial radii, Simionescu et al.\ (2017) measure a hydrostatic mass of  $M_{\rm 200}=1.05\times10^{14}$\,M$_{\odot}$ and $R_{\rm vir}\approx R_{200}=974$\,kpc.
Its asymmetric X-ray morphology, multiple subclumps, and low mass all suggest that Virgo is a dynamically young cluster that is in the epoch of active accretion. 


The M49 group resides at $r=1.25\,R_{\rm vir}$ (4$^{\circ}$ from the cluster center) and is falling onto the southern outskirts of Virgo (Figure~\ref{fig:suzaku}). Its brightest group galaxy (BGG), also called M49 (NGC~4472, $M_{\rm K}=-25.78$), is optically brighter than M87 ($M_{\rm K}=-25.38$), the central galaxy of Virgo (ATLAS$^{\rm 3D}$, Cappellari et al.\ 2011). Throughout this paper ``M49" designates the M49 group rather than the BGG alone. 
The enriched X-ray atmosphere of M49 was first detected by the {\sl ROSAT} X-ray observatory (Forman et al.\ 1993). 
Previous X-ray studies of M49 were confined to the central 10--50\,kpc.
M49 is a cool core system with an average temperature of 1.3\,keV.
It harbors two X-ray cavities at a radius of $\approx4$\,kpc filled with radio emission, demonstrating  feedback from its active galactic nucleus (AGN) (Biller et al.\ 2004; Gendron-Marsolais et al.\ 2017).
A prominent surface brightness edge is present $\sim20$\,kpc north of the group center ({\sl ROSAT}--Irwin \& Sarazin 1996; {\sl Chandra}--Biller et al.\ 2004; {\sl XMM-Newton}--Kraft et al.\ 2011), implying motion towards M87. 

M49 provides the best opportunity to study accretion at cluster outskirts. 
Thanks to its proximity and brightness, we can study features with a sensitivity and linear resolution unachievable in any other object.
In particular, the spatial resolution of observations matches that of high resolution simulations, bridging the gap between the microscale cluster physics and the macroscale cluster evolution. 
The radial velocity of M49 (958\,km\,s$^{-1}$) is close to the median value of the Virgo cluster (1088\,km\,s$^{-1}$) (Mei et al.\ 2007),
implying that M49 is moving close to the plane of the sky, which minimizes projection effects. {\sl XMM-Newton} is particularly suitable to study the gas properties of galaxy groups, thanks to its relatively low and stable background at $\lesssim$\,1.5\,keV, quality resolution, and large effective area. 
In this paper, we present the group gas properties of M49 out to radii $>150$\,kpc in multiple directions using mosaic {\sl XMM-Newton} observations and use them to probe the entire gas dynamics and merger history of M49.


We adopt a luminosity distance of 16.7\,Mpc for M49 (Blakeslee et al.\ 2009). This corresponds to an angular scale of $1^{\prime\prime}=0.081$\,kpc and a redshift of 0.0038 for a cosmology with $H_0=69.6$\,km\,s$^{-1}$\,Mpc$^{-1}$, $\Omega_{\Lambda}=0.714$, and $\Omega_m=0.286$. 
We describe the observations and data reduction in \S2. We report the thermal and chemical properties and mass distribution of M49 in \S3.
The implications for the dynamics and merging history of this group are discussed in \S4, and our conclusions are summarized in \S5. Uncertainties reported in this paper are at 1$\sigma$ unless stated otherwise.

\section{\bf observations and data reduction}

\subsection{XMM-Newton}
Our analysis includes 9 {\sl XMM-Newton} pointings within a 1$^{\circ}$ radius of M49: 5 central pointings (C and N) and 4 offset pointings (W, SW, and SE), in total $\sim500$\,ksec (unfiltered) exposure time as indicated in Figure~\ref{fig:suzaku} and listed in Table~1. 
Data analysis was performed using the {\sl XMM-Newton} Science Analysis System (SAS) version xmmsas-v15.0.0. 
All the ODF files were processed using {\tt emchain} and {\tt epchain} to ensure the latest calibrations. Soft flares were filtered from MOS and pn data using the XMM-ESAS tools {\tt mos-filter} and {\tt pn-filter} respectively (Snowden \& Kuntz 2011). 
The effective exposure time of each detector is listed in Table~1. In all but two observations, the MOS1 CCDs 3 and 6 are lost to micro-meteorites. 
We only include events with ${\rm FLAG} = 0$ and ${\rm PATTERN}\le12$ for MOS data and with ${\rm FLAG} = 0$ and ${\rm PATTERN}\le4$ for pn data.    
Point sources detected by the XMM-ESAS routine {\tt cheese} and confirmed by eye were excluded from further analysis. 
Corrections were applied in spectral and imaging analysis to remove the effects of out of time pn
events due the X-ray bright core of M49.

 \begin{deluxetable*}{ccccccc}

\tablewidth{0pc}
 \centering
\tablecaption{Observational log of the M49 group}
\tablehead{
\colhead{Obs ID}&\colhead{Name}&\colhead{Obs-Date}&\colhead{Exposure (ksec)$^*$}&\colhead{RA (J2000)}&\colhead{Dec (J2000)}&\colhead{PI}}
\startdata
\multicolumn{7}{c}{XMM-Newton} \\
\hline
0112550601&M49-N&2002-06-05&25 (13, 13, 9)&12 29 46.00&+07 59 47.0&Turner\\
0200130101&M49-N&2004-01-01&111 (79, 78, 68)&12 29 46.75&+07 59 59.9&Maccarone\\
0722670101&M49-W &2013-12-28&73 (60, 61, 38)&12 27 57.67&+07 54 30.0&Kraft\\
0722670201&M49-SW &2013-12-30&73 (54, 59, 41)&12 28 44.18&+07 41 18.8&Kraft\\
0722670301&M49-SE &2014-01-09&73 (20, 21, 9)&{12 30 15.30}&{+07 41 30.0}&Kraft\\
0722670601&M49-SE &2014-01-13&32 (9, 10, 3)&12 30 15.30&+07 41 30.0&Kraft\\
0761630101&M49-C&2016-01-05&118 (88, 88, 70)&{12 29 39.70}&{+07 53 33.0}&Maccarone\\
0761630201&M49-C&2016-01-07&118 (82, 81, 64)&12 29 39.70&+07 53 33.0&Maccarone\\
0761630301&M49-C&2016-01-09&117 (82, 82, 68)&12 29 39.70&+07 53 33.0&Maccarone\\
\hline
\multicolumn{7}{c}{Suzaku} \\
\hline
801064010&M49&2006-12-03&121 (97, 99, 100)&12 29 46.58&+08 00 18.0&Loewenstein\\
807122010&VIRGO S8&2012-07-05&16 (14, 13, 14)&12 30 21.26&+09 54 57.6&Simionescu\\
807123010&VIRGO S9&2012-07-07 &19 (17, 17, 17)&12 30 16.73&+09 35 19.7&Simionescu\\
807124010&VIRGO S10&2012-12-08&16 (12, 12, 12) &	12 30 12.96&+09 16 24.6&Simionescu\\
807125010&VIRGO S11&2012-12-08&17 (17, 17, 17) &12 30 07.80&+08 56 35.5&Simionescu\\
807126010&VIRGO S12&2012-12-09 &16 (14, 14, 14) &12 29 58.87&+08 37 08.4&Simionescu\\
807127010&VIRGO S13&2012-12-09&13 (10, 10, 10) &12 29 54.19&+08 18 43.2&Simionescu\\
807128010&VIRGO S14&2012-12-09&15 (15, 15, 15)&12 29 45.02&+07 40 56.3&Simionescu\\
807129010&VIRGO S15&2012-12-10&18 (15, 15, 15)&12 29 35.66&+07 21 17.3&Simionescu\\
807130010&VIRGO S16&2012-12-10&22 (20, 20, 19)&12 29 31.51&+07 03 11.2&Simionescu\\
807131010&VIRGO S17&2012-12-11&19 (17, 16, 17)&12 29 26.74&+06 44 42.7&Simionescu\\
807132010&VIRGO S18&2012-12-11 &20 (17, 17, 17)&12 29 21.91&+06 25 14.2&Simionescu\\
807133010&VIRGO S19&2012-12-11&19 (19, 18, 18)&12 29 17.76&+06 07 05.5&Simionescu
\enddata
\tablecomments{*Effective exposure times of MOS1, MOS2, and pn for {\sl XMM-Newton} and XIS0, XIS1, and XIS3 for {\sl Suzaku} are listed in the brackets. Quadrant 0 pn event of 0722670301 displays abnormalities and is not included in the analysis.}
\end{deluxetable*}

\subsubsection{\it Background modeling}

We consider two sources of background components: Astrophysical X-ray background (AXB) and Non-X-ray background (NXB).  
The AXB model contains a power law model {\tt pow}$_{\rm CXB}$ with a photon index fixed at $\Gamma=1.46$ characterizing the cosmic X-ray background (CXB, De Luca \& Molendi 2004), a thermal emission model {\tt apec}$_{\rm MW}$ with a temperature allowed to vary between 0.15\,keV and 0.6\,keV representing the Milky Way emission (McCammon et al.\ 2002), and another thermal emission model  {\tt apec}$_{\rm LB}$ with a temperature fixed at 0.11\,keV for the Local Bubble emission. Metal abundance and redshift were fixed at 1 and 0 respectively for {\tt apec}$_{\rm MW}$ and {\tt apec}$_{\rm LB}$. All the components except {\tt apec}$_{\rm LB}$ are expected to be absorbed by foreground (Galactic) cooler gas, characterized by the {\tt phabs} model. 
The AXB components were determined via a joint-fit of offset pointings (W, SW, SE) and a RASS spectrum extracted from a region farther south of M49.
The best-fit parameters are listed in Table~2. 

The NXB model contains a set of fluorescent instrumental lines and a continuum spectrum for each MOS and pn detector. 
Fluorescent instrumental lines produced by the hard particles are modeled with a set of gaussian lines. Their centroid energies are listed in Snowden \& Kuntz (2011) and we set an upper limit of 0.3\,keV on each line width. We use a broken powerlaw model, {\tt bknpow}, to characterize the continuum particle background component. The energy break is fixed at 3\,keV. 
To constrain the NXB components, we obtain filter wheel closed (FWC) observations taken at the nearest possible time to each observation. 
For each region of interest, we simultaneously fit the M49 and the FWC data with their NXB parameters linked (ratios of their {\tt bknpow} normalizations are determined by simultaneously fitting spectra extracted from the unexposed corners of both the FWC and the M49 observations).
An additional quiescent level of soft proton flares may affect some detectors. 
We compare the area-corrected count rates in the 6--12\,keV energy band within the field of view (excluding the central 10$^{\prime}$) and that in the unexposed corners of the detector. 
If this ratio is below 1.15, we consider the observation not contaminated by the residual soft proton flare (Molendi et al.\ 2004). For contaminated observations, we add a power law {\tt pow}$_{\rm SB}$ to model this residual soft proton component. Its photon index is allowed to vary between 0.1 and 1.4 (Snowden \& Kuntz 2011). 


 \begin{deluxetable}{ccccc}

\tablewidth{0pc}
 \centering
\tablecaption{X-ray Background model parameters}
\tablehead{
\colhead{Background}&\colhead{kT or $\Gamma$}&\colhead{XMM norm$^{*}$}&\colhead{kT or $\Gamma$}&\colhead{Suz norm$^{*}$}}
\startdata
CXB&1.46&$0.88\pm0.04$&1.5&0.89\\
MW&$0.183\pm0.001$&$1.97\pm0.06$&0.2&2.2\\
LB&0.11&$1.17\pm0.04$&0.104&1.46
\enddata
\tablecomments{Model parameters for {\sl XMM-Newton} observations are determined in this work; those for {\sl Suzaku} observations are taken from Simionescu et al.\ (2017).  * $\times10^{-6}$ normalization per arcmin$^2$. }
\end{deluxetable}

\subsubsection{\it Spectral and imaging analyses}

Spectral analysis was restricted to the 0.5--10.0 keV and 0.7--10.0 keV energy bands for MOS and pn, respectively. Each observation and its corresponding FWC data were fit jointly to two sets of models.
The first model set takes the form of {\tt phabs}$\times$({\tt pow}$_{\rm CXB}$+{\tt apec}$_{\rm MW}$+{\tt vapec}$_{\rm group}$)+{\tt apec}$_{\rm LB}$. Components of this model set were fixed at zero for the FWC data. 
Parameters of the AXB models were fixed to the values determined with offset pointings and listed in Table~2.
The thermal {\tt vapec}$_{\rm group}$ model is for the M49 emission. 
Abundances of Si, S, and Fe were allowed to vary freely;  
O, Ne, Mg, Ar, Ca, and Ni were associated to Fe based on the best-fits and uncertainties of the abundance ratios of 44 clusters, groups, and massive ellipticals (the CHEERS sample) presented in Mernier et al.\ (2016). 
The solar abundance standard of Asplund et al.\ (2009) was adopted. The measured Fe abundance needs to be multiplied by 1.07 and 0.68 when compared to Lodders (2003) and Anders \& Grevesse (1989), respectively. 
We included an additional power law component with an index of 1.6 for regions at the BGG center to model the unresolved emission from low mass X-ray binaries (LMXB, Irwin et al.\ 2003). 
The second model set characterizes the NXB components and the spectra were not folded through the Auxiliary Response Files (ARF). 
Parameters of the NXB models were linked between the observation and the FWC data. 
The spectral fit was performed using XSPEC 12.7.0 (Arnaud 1996) and $\chi^2$-statistics. We adopted a Galactic hydrogen column of $N_H =1.6\times10^{20}$\,cm$^{-2}$ toward M49, determined from the Dickey \& Lockman (1990) map. 


\begin{figure}[h]
   \centering
    \includegraphics[width=0.5\textwidth]{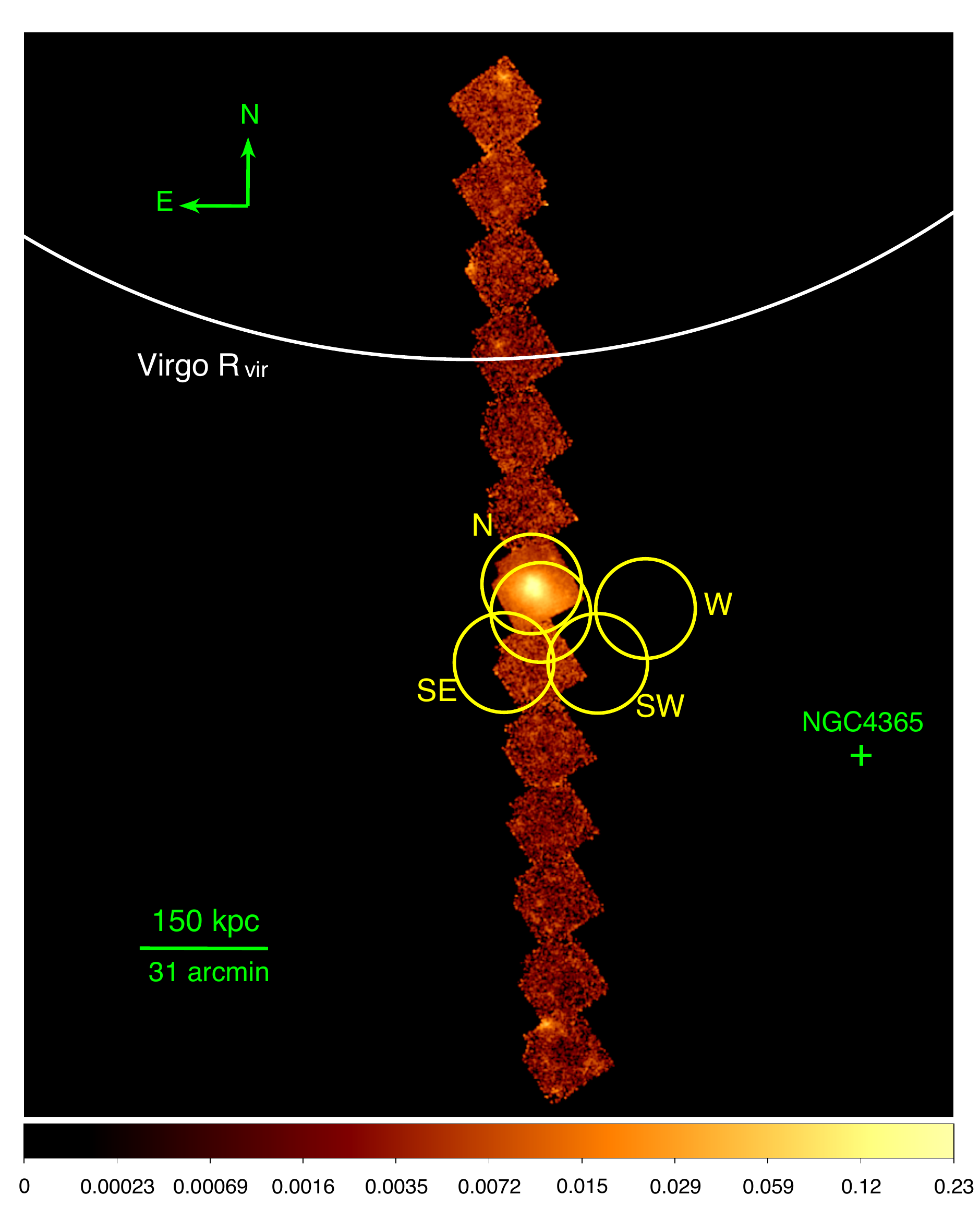}
\figcaption{\label{fig:suzaku} Mosaic {\sl Suzaku} image of the M49 group and its surrounding in the 0.5--2.0 keV energy band in units of cts/s/pixel. The image is exposure and vignetting corrected with Non-X-ray background subtracted. The white curve marks the virial radius ($R_{\rm vir}\approx R_{200}=974$\,kpc) of the Virgo cluster. Yellow circles indicate the field-of-view of the {\sl XMM-Newton} observations used in this study. Green cross marks the location of the W' cloud at a distance of 23\,Mpc and centered on NGC~4365.}
\end{figure}

\begin{figure}[h]
   \centering
    \includegraphics[width=0.5\textwidth]{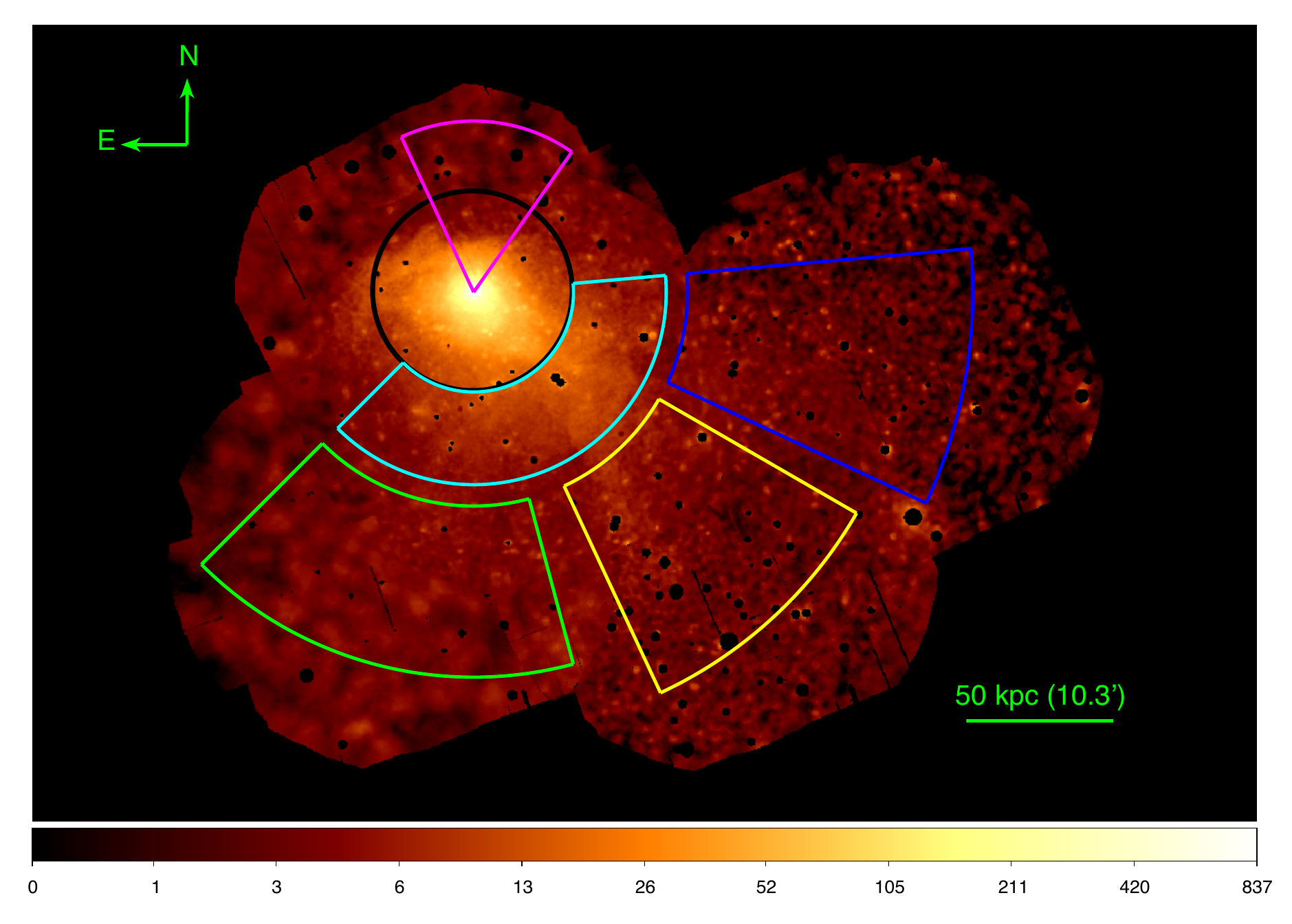}
\figcaption{\label{fig:xmm} Mosaic {\sl XMM-Newton} image of the M49 group in the 0.5--2.0 keV energy band and in units of cts/s/deg$^{2}$. The image is exposure and vignetting corrected with instrumental background subtracted. The colored regions are defined for the spatial spectroscopic analysis whose results are plotted in Figures 4, 6, and 7.}
\end{figure}

We produce background-subtracted, vignetting and exposure corrected EPIC mosaic images of M49. Individual detector images were created using the XMM-ESAS tasks
{\tt mos-spectra} and {\tt pn-spectra}, and combined with the {\tt comb}
routine. We used the {\tt adapt\_900} routine to bin each image in $2\times2$ pixels and adaptively smooth it with a minimum of 50 counts. The resulting image is shown in Figure~\ref{fig:xmm}. 

\begin{figure}[h]
   \centering
       \includegraphics[width=0.475\textwidth]{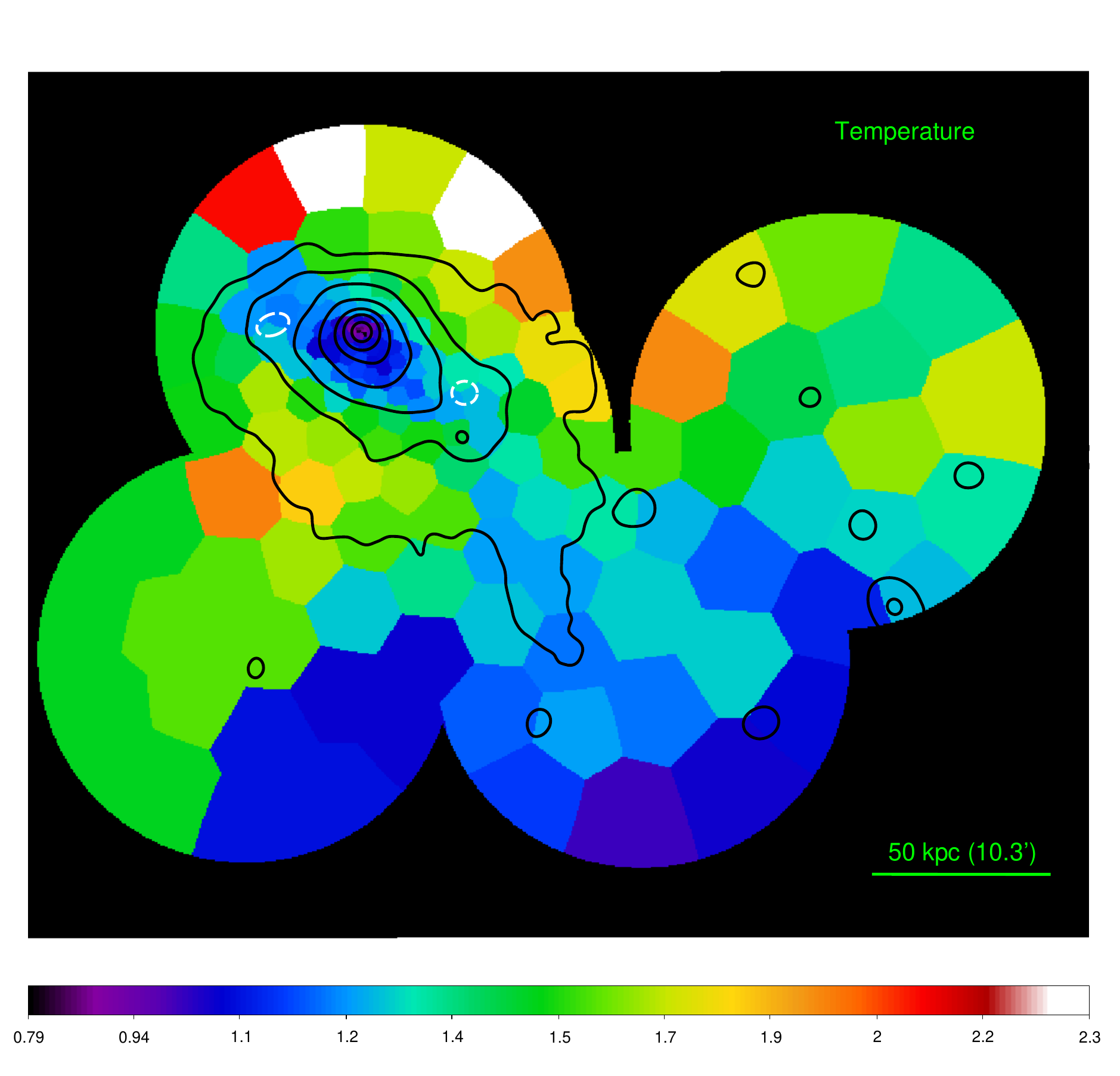}
    \includegraphics[width=0.475\textwidth]{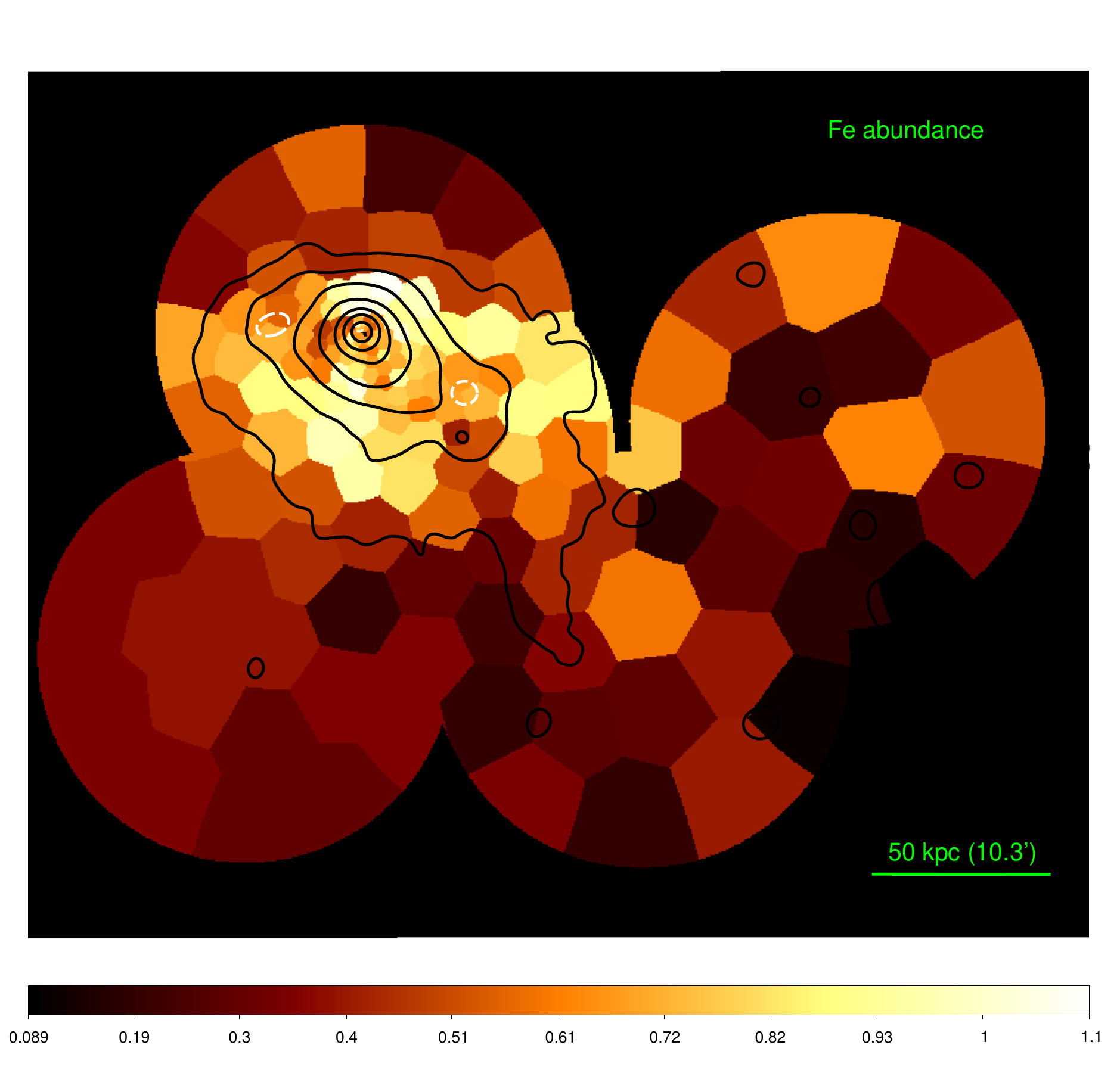}
\figcaption{\label{fig:map} {\sl XMM-Newton} Spectroscopic maps with the X-ray contour in the 0.7--1.3 keV energy band overlaid. {\it top:} Temperature map of M49 in units of keV.  {\it bottom:} Fe abundance map of M49 in units of the solar abundance derived with a single temperature model. White circles indicate the location of ghost cavities.}
\end{figure}

We performed a two-dimensional spectral analysis using Weighted Voronoi Tesselation (WVT) binning (Diehl \& Statler 2006)
based on the Voroni binning algorithm presented in Cappellari \& Copin (2003). We generated a WVT binning image containing 173 regions for the image in the 0.5-2.0 keV energy band with a S$/$N of at least 80 in each bin. 
We used a single {\tt vapec} model to probe the hot gas emission in each region. 
The resulting temperature and Fe abundance maps are shown in Figure~\ref{fig:map}.

\subsection{Suzaku}
To complement the {\sl XMM-Newton} analysis, 
we reduced all the existing {\sl Suzaku} observations within 2$^{\circ}$ of M49. The observations were taken along the north-south directions, including 12 pointings as part of a {\sl Suzaku} Key Project to map the Virgo cluster out to $R_{\rm vir}$ in four directions (Simionescu et al.\ 2015, 2017) and one deep ($\sim$100\,ksec) observation centered on M49. The observational log is listed in Table~1. 
We produce a mosaic image that is exposure and vignetting corrected with NXB subtracted as shown in Figure~\ref{fig:suzaku}.
Data reduction and analysis were performed with {\sl HEASOFT6.19} using CalDB20160607. Detailed procedures are stated in Su et al.\ (2013) and Su et al.\ (2015). In brief, all data were converted to the $3\times3$ modes. Events with a low geomagnetic cutoff rigidity (COR $<6$ GV) and an Earth elevation $<10^{\circ}$ were removed. The calibration source and hot pixels were excluded. Flares exceeding the average count rate by $3\sigma$ were filtered. Bright sources were identified by eye. A circular region with a radius of $2\arcmin$ centered on each source was excluded. 
We extracted spectra from a circular region with a radius of $7^{\prime}$ from each pointing, which is sufficiently broad for the point-spread function (PSF) of {\sl Suzaku}. 
The FTOOL
\texttt{xissimarfgen} was used to generate an ARF for each region
and detector
assuming uniform sky emission with a radius of $20^{\prime}$.
Redistribution matrix files (RMF) and NXB spectra were generated for each region
and detector with \texttt{xisrmfgen} and
\texttt{xisnxbgen} respectively. Spectra extracted from all pointings were fit simultaneously; parameters of the outermost four pointings were linked. We left out the pointing centered on the BGG M49 since it is much deeper and brighter in X-rays than all other regions and we do not wish to let it dominate the simultaneous fit. 
Energy ranges were restricted to 0.5--7.0 keV for
the back-illuminated CCD (XIS1) and 0.6--7.0 keV for the
front-illuminated CCDs (XIS0, XIS3). We fit each spectrum to the model ${\tt phabs}\times({\tt vapec}_{\rm ICM}+{\tt pow}_{\rm CXB}+{\tt apec}_{\rm MW})+{\tt apec}_{\rm LB}$. 
The model parameters of the AXB components were fixed to the values determined for the southern side of Virgo by Simionescu et al.\ (2017) and listed in Table~2.

\section{\bf Results}

\begin{figure*}[h]
   \centering
    \includegraphics[width=0.95\textwidth]{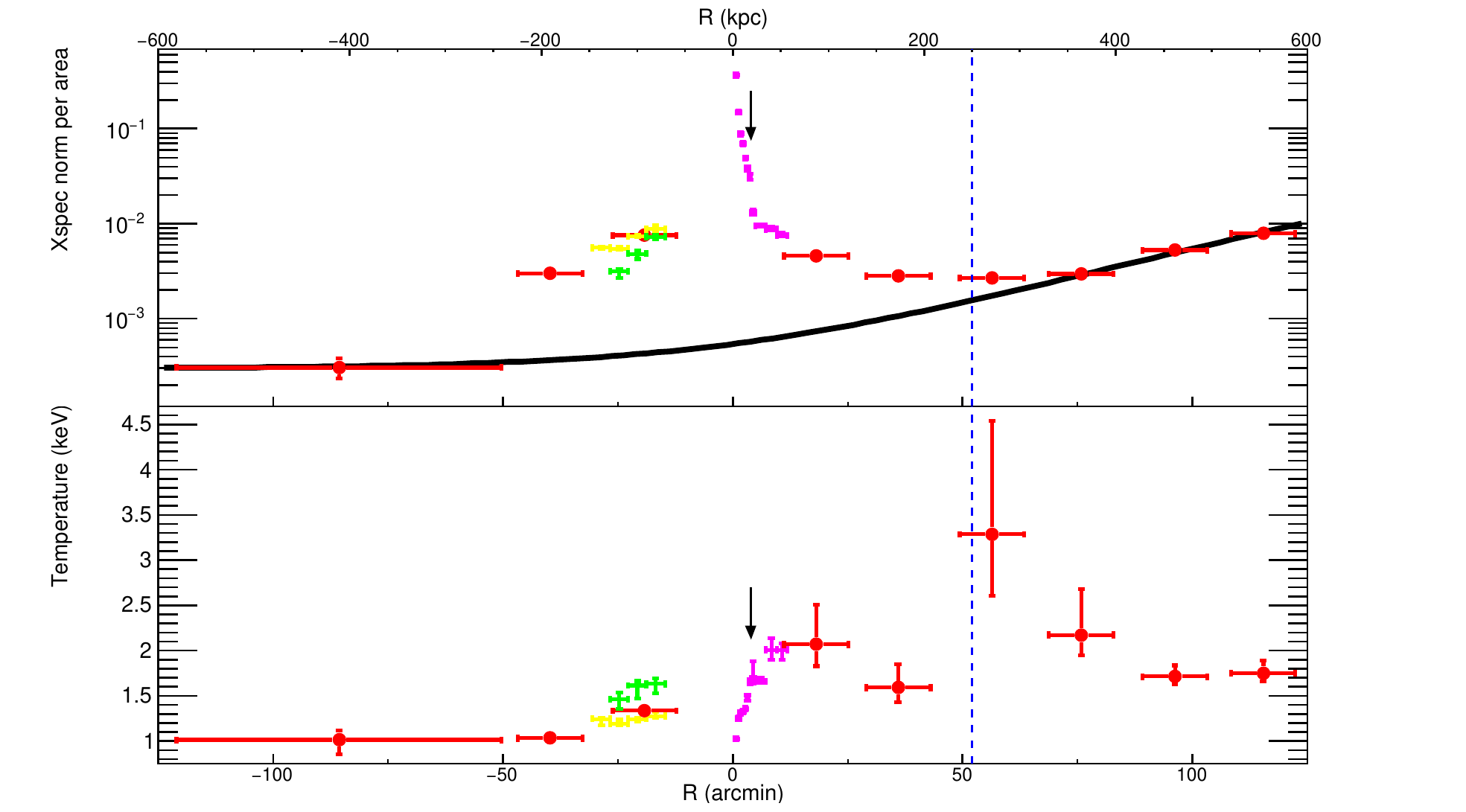}
\figcaption{\label{fig:norm} Radial profiles of the projected spectrum normalization per unit area (a circular region with a radius of $20^{\prime}$) and the hot gas temperature centered on M49 with $x$-axis increasing northward. Red circles: results from {\sl Suzaku}. 
Data of other colors are derived with {\sl XMM-Newton} observations. The corresponding regions are indicated in Figure~\ref{fig:xmm}. Black solid line is the best-fit normalization profile in the absence of M49, indicating the underlying Virgo ICM distribution. The vertical arrow marks the northern surface brightness discontinuity. The vertical blue line marks the virial radius (R$_{200}$) of Virgo.}
\end{figure*}

We present the radial profiles of 
the normalization (per area) and temperature of the thermal emission model, {\tt vapec}$_{\rm ICM}$, to the north and south of M49 reaching a radius of 500\,kpc (Figure~\ref{fig:norm}), derived from the {\sl Suzaku} observations. The normalization profile displays a hump-shaped feature on top of a gradually-declining profile, demonstrating the presence of the M49 group in the outskirts of the Virgo ICM. The group gas reaches radii of 250\,kpc and 400\,kpc to the north and south, respectively. 
M49 is a BGG with its own extended atmosphere, as opposed to a bright
early-type galaxy, and its atmosphere has come into direct contact
with the Virgo ICM.
Below we present the gas properties of this infalling group.

\subsection{\bf Surface brightness and contact discontinuity}

\begin{figure}[h]
   \centering
    \includegraphics[width=0.5\textwidth]{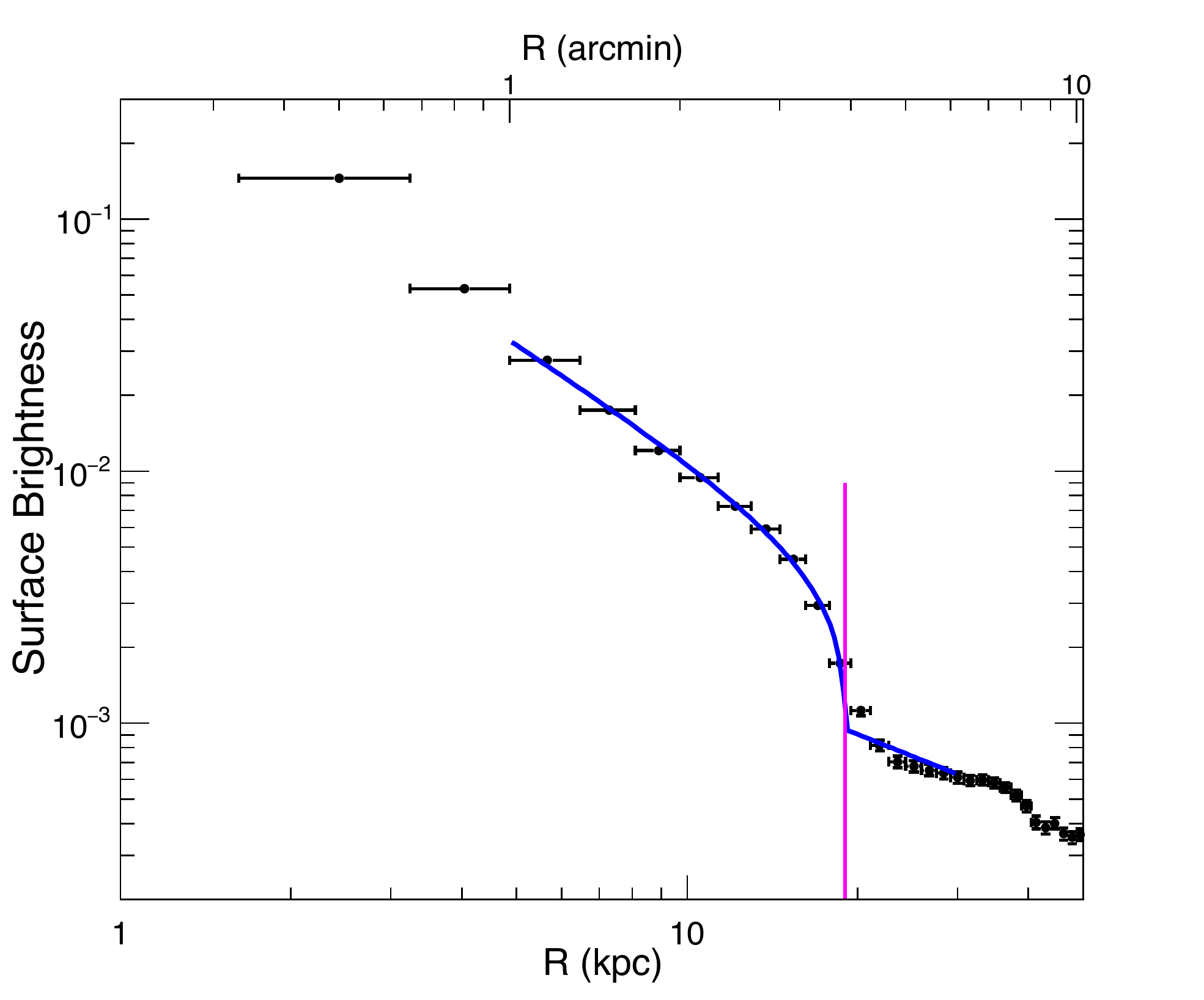}
\figcaption{\label{fig:sur} Surface brightness profiles of the upstream region derived with {\sl XMM-Newton} observations in the energy band of 0.7--1.3 keV and in units of cts/s/arcmin$^2$. The profile is centered on M49. Blue solid line: the best-fit of a broken powerlaw density model. The bright surface brightness edge is marked by the solid vertical magenta line.} 
\end{figure}

In Figure~\ref{fig:sur}, we present the {\sl XMM-Newton} surface brightness profile of M49 in the 0.7--1.3 keV energy band to the north. The energy band was chosen to maximize the source-to-background ratio (Ettori et al.\ 2010; Kraft et al.\ 2017).
The previously reported bright edge (Irwin \& Sarazin 1996; Biller et al.\ 2004; Kraft et al.\ 2011) is visible at $\sim20$\,kpc north of the group center (marked by the magenta solid line). South of this edge may be the regime of the interstellar medium (ISM) of the BGG M49. This is consistent with the strong gradient in metallicity across this edge (see \S3.2). 
We fit the surface brightness profile across this edge to a broken power-law density model.
We obtain a break at $19.0\pm0.1$\,kpc from the group center. 
Taking into account the difference in the cooling function on each side, we obtain a best-fit density jump of $1.9\pm0.3$.

\subsection{\bf Thermal and chemical distributions}

\begin{figure}[h]
   \centering
       \includegraphics[width=0.5\textwidth]{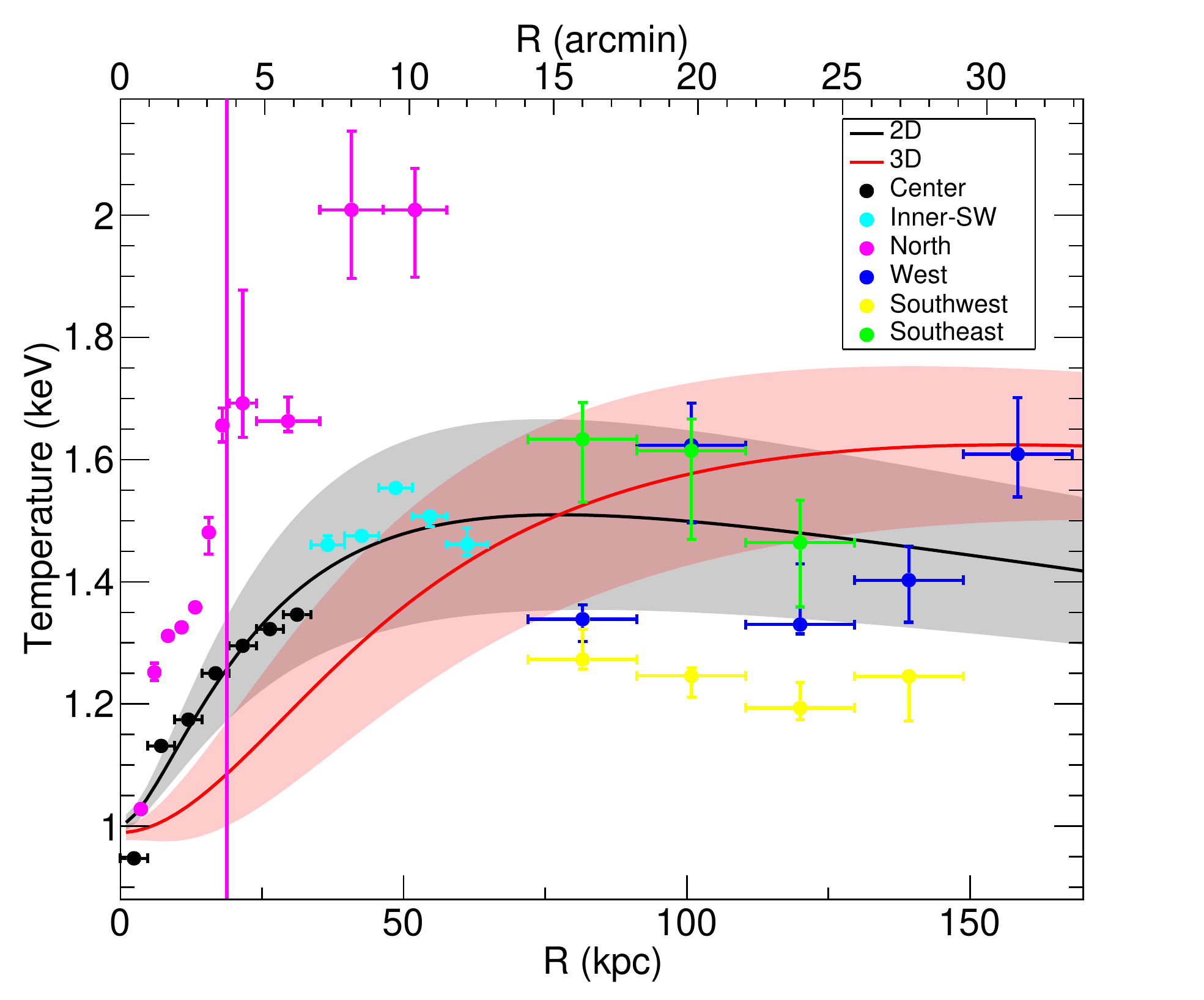}

\figcaption{\label{fig:west} Projected temperature profiles centered on M49 derived with {\sl XMM-Newton}. The corresponding regions are indicated in Figure~\ref{fig:xmm}. The best-fit 2D and 3D profiles of the central--western regions (black, cyan, and blue) are plotted in the black and red solid lines, respectively. The vertical magenta solid line marks the northern edge in surface brightness}
\end{figure}

\begin{figure}[h]
   \centering
       \includegraphics[width=0.5\textwidth]{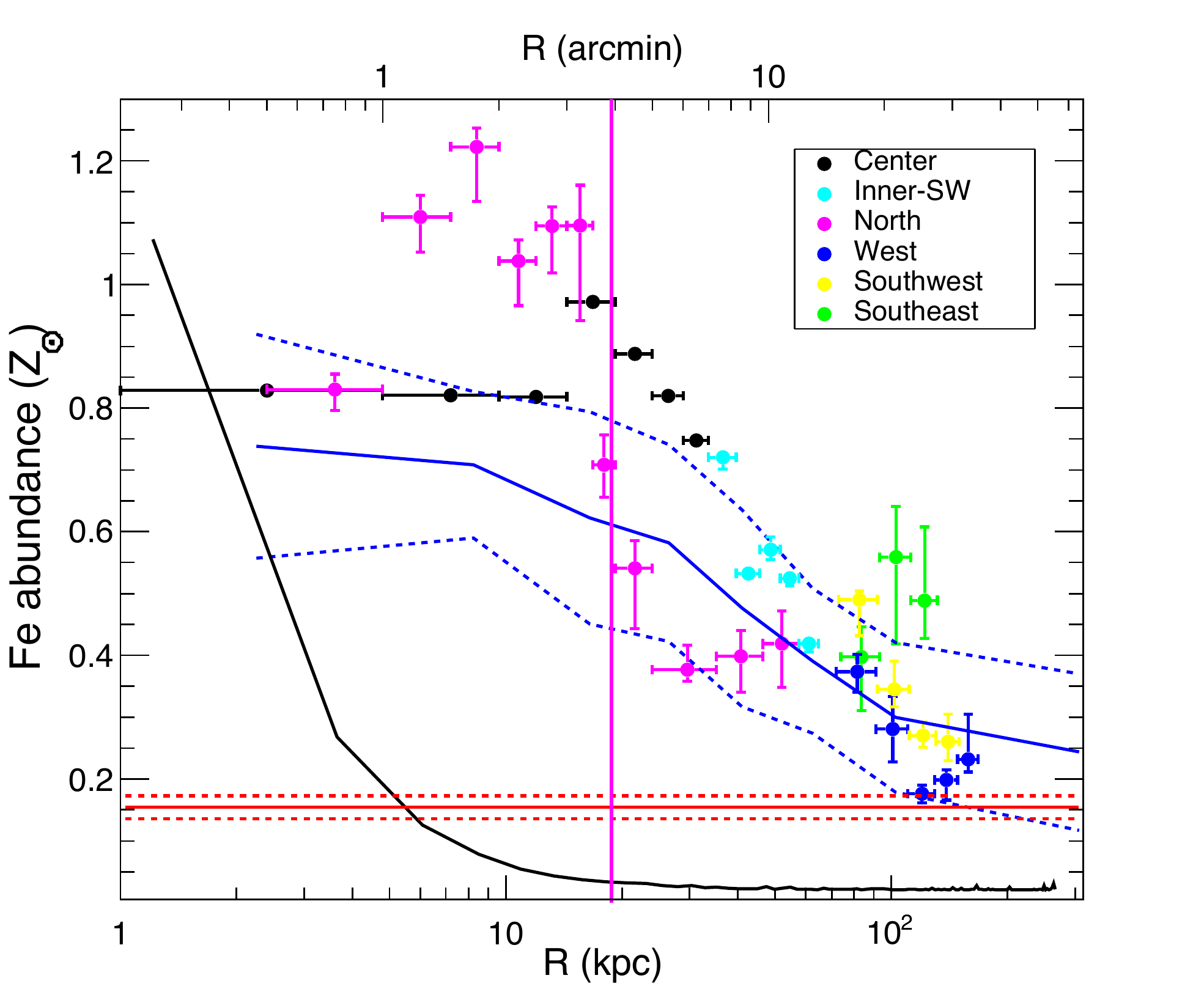}
\figcaption{\label{fig:abun} Metallicity profiles centered on M49 derived with {\sl XMM-Newton}. The corresponding regions are indicated in Figure~\ref{fig:xmm}. Red lines: the range of the Virgo ICM Fe metallicity on its southern outskirts taken from Simionescu et al.\ (2015). Blue lines: the range of the average Fe metallicity profile of typical galaxy groups taken from Mernier et al.\ (2017). Black solid line: the scaled $K$-band surface brightness profile of M49. The vertical magenta solid line marks the northern edge in surface brightness.}
\end{figure}

In Figures~\ref{fig:west} and \ref{fig:abun}, we present radial profiles of temperature and Fe abundance centered on M49 in 4 directions, derived from the {\sl XMM-Newton} observations. 
Different extraction regions are coded in the colors indicated in Figure~\ref{fig:xmm}. The temperature is below 1 keV at the group center and gradually increases to $\sim1.5$\,keV at $r=50$\,kpc and stays between 1\,keV\,$\sim$\,2\,keV to a radius of 160 kpc. 
We note an increase in temperature on the fainter side of the northern edge, implying it is a cold front (Figure~\ref{fig:west}). 
The {\sl XMM} image in Figure~\ref{fig:tail} reveals two sets of cool filamentary arms to the east and west of the nucleus at radii of 25--35\,kpc, possibly associated with X-ray cavities produced in a previous AGN outburst; the lack of radio counterparts makes them ``ghost cavities" (also see Kraft et al.\ 2011). 
The temperature map (Figure~\ref{fig:map}--top) reveals a 
channel filled with cooler gas 
elongated east-west out to a radius of $\lesssim50$\,kpc. 
This is similar to the thermal structure of NGC~1399, the central galaxy of the Fornax cluster (Su et al.\ 2017c). 
In these systems, cool gas from the cluster center may have been lifted up by buoyant bubbles. {We note that the northern edge of the BGG is remarkably sharp, but with a bump feature extending from its eastern side (Figures~\ref{fig:xmm} and \ref{fig:tail}). This feature is likely to be low entropy gas lifted up by the eastern bubble (Figure~\ref{fig:map}--top). Being denser than the group gas at this radius would make the cool gas slower to
respond to the ram pressure, disrupting the cold front.} 
The original cold front may be facing northeast rather than due north. In other words, M49 may be heading northeast, which naturally explains the stripped tail to the southwest (\S3.3).


The metallicity profiles outside the core decline from $\lesssim1$\,Z$_{\odot}$ to $\gtrsim0.2$\,Z$_{\odot}$ over a radial range of $>100$\,kpc and stay above the metal abundance of the Virgo outskirts (Simionescu et al.\ 2015). The metallicity profile of M49 is broadly consistent with that of typical galaxy groups presented in Mernier et al.\ (2017). In Figure~\ref{fig:abun}, we plot the scaled $K$-band surface brightness profile derived from the {\sl WISE} infrared image. The stellar component of M49 reaches the background level outside a radius of 10\,kpc. The distribution of the hot gas that is more enriched than the Virgo ICM 
extends far beyond the domain of the BGG, corroborating the hypothesis that M49 possesses its own group gas.  
The temperature and metallicity profiles to the north change more quickly than in other directions, implying that the BGG has moved northward relative to the group gas.

A drop in metal abundance is observed at the center of M49 (Figure~\ref{fig:abun}). Such a central metallicity drop has been widely observed in clusters, groups and elliptical galaxies (Sanders et al.\ 2016; Mernier et al.\ 2017). {The deposition of metals into dust grains may be responsible for the metallicity drop in some systems 
(Lakhchaura et al. 2019; Panagoulia et al.\ 2013; 2015), whereas M49 is nearly dust free (Temi et al.\ 2007).} 
The metallicity map (Figure~\ref{fig:map}--bottom) reveals that the distribution of the central metallicity drop of M49 deviates from the azimuthal symmetry. The low metallicity gas is coincident with the cool gas elongated east-west out to the ghost cavities at $r\lesssim50$\,kpc. 
A similar coincidence between a low metallicity and the radio lobes at $r\sim5$\,kpc has been noted by the {\sl Chandra} study of Gendron-Marsolais et al.\ (2017). 
The central low metallicity and cool gas may have been uplifted by AGN outbursts over multiple cycles.

\subsection{\bf The stripped tail}

\begin{figure*}[h]
   \centering
       \includegraphics[width=1.0\textwidth]{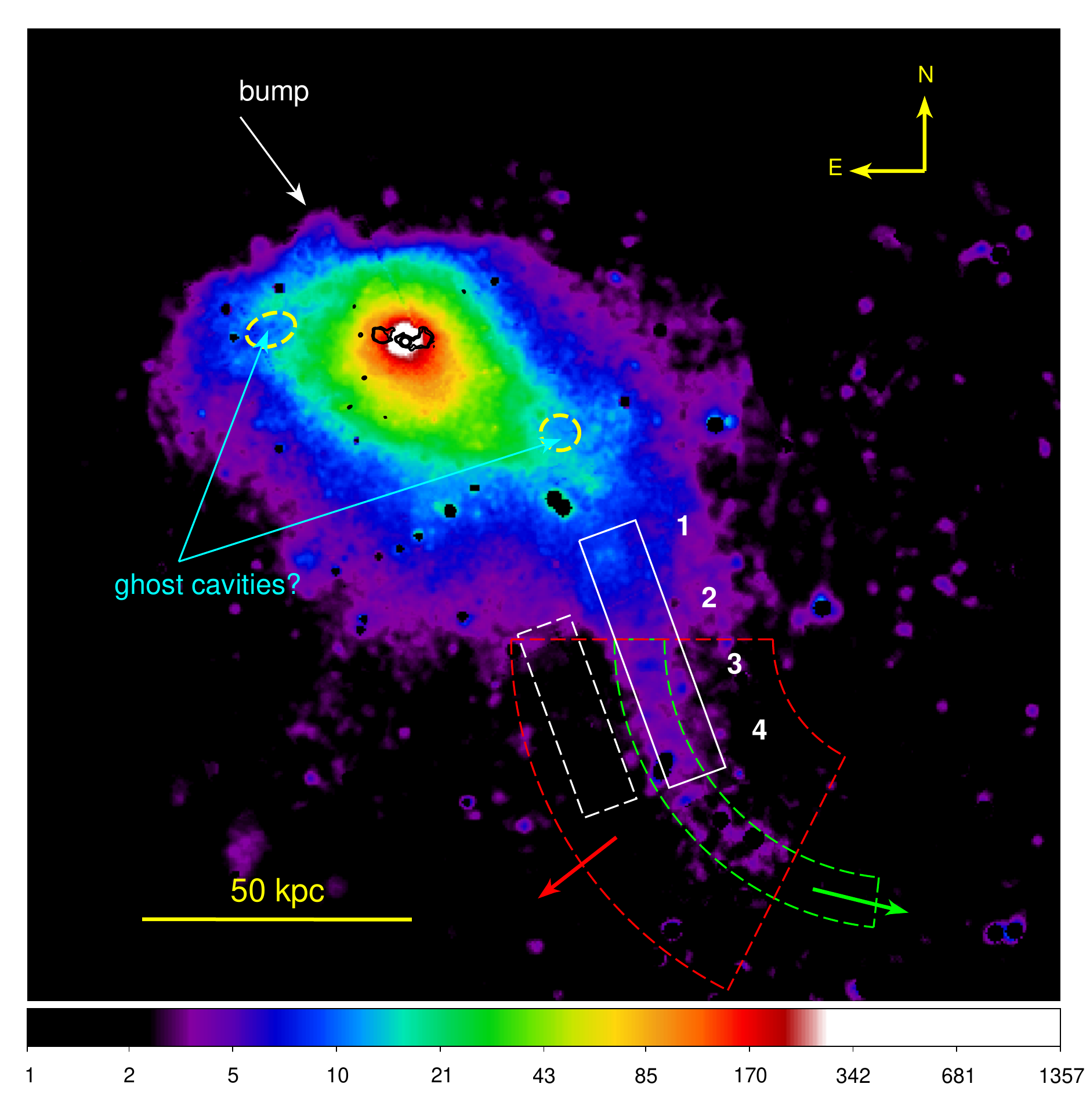}

\figcaption{\label{fig:tail} {\sl XMM-Newton} image in the 0.7--1.3 keV energy band. A stripped tail is visible to the southwest of the BGG. We extract spectra from the white solid box for the tail gas and the white dashed box for the ambient group gas. Black contour: radio emission from the VLA First survey. Two ghost cavities are marked by dashed yellow circles. We derive surface brightness profiles along and across the tail using the green and red annular sectors.}
\end{figure*}

\begin{figure}[h]
   \centering
       \includegraphics[width=0.5\textwidth]{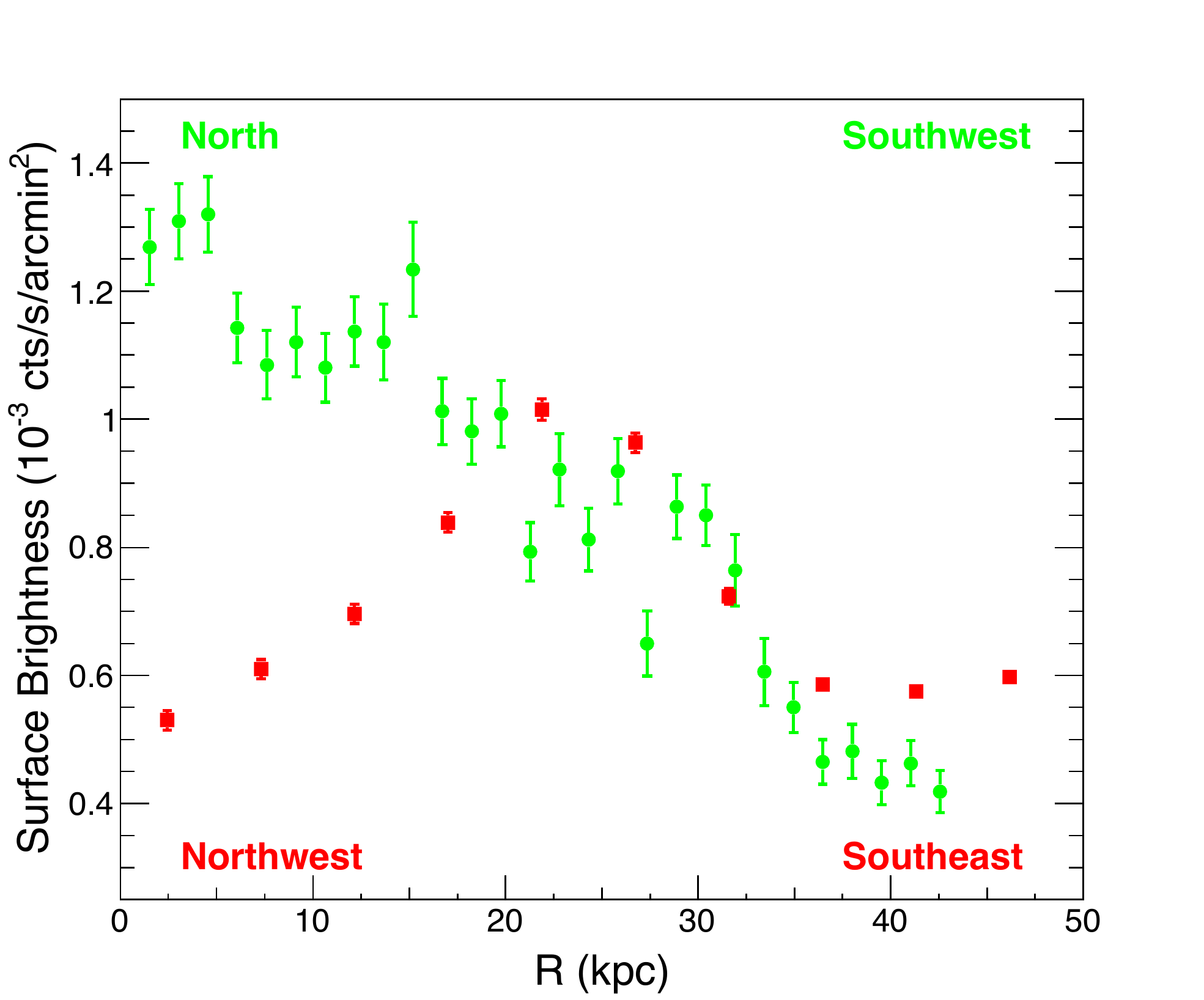}
\figcaption{\label{fig:tail_sur} Surface brightness profiles along and across part of the southwest tail in the energy band of 0.7--1.3 keV. The circle and square data points correspond to the green (N--SW) and red (NW--SE) annular sectors, respectively, marked in Figure~\ref{fig:tail}.}
\end{figure}

\begin{figure}[h]
   \centering
       \includegraphics[width=0.5\textwidth]{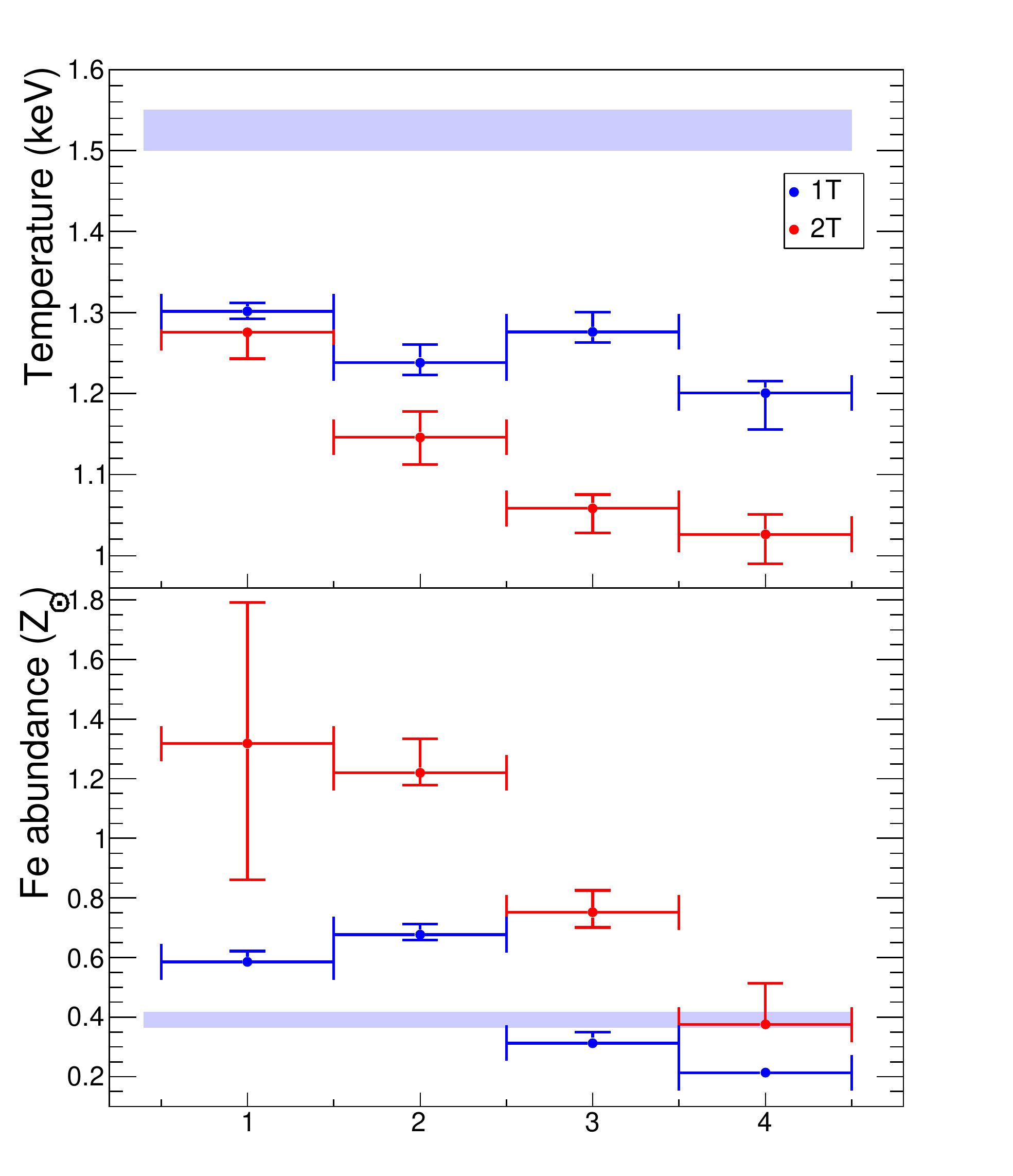}
\figcaption{\label{fig:box} Gas properties of the stripped tail compared to the ambient group gas. Temperatures and Fe abundances measured along the tail (corresponding to the solid box region in Figure~\ref{fig:tail}). Blue data points: results of a single-temperature model. Red data points: results of a two-temperature model; parameters of the hotter temperature component is fixed at the best-fit of the ambient group gas (light blue lines, corresponding to the dashed box region in Figure~\ref{fig:tail}). Red data points reflect the intrinsic temperatures and Fe metallicities of the tail gas.}
\end{figure}

The group center, dominated by the ISM of M49, is significantly brighter in X-rays than the ambient atmosphere. As shown in Figure~\ref{fig:tail}, we detect a 70\,kpc long and 10\,kpc wide gaseous tail adjacent to the BGG to the southwest, nearly opposite to the contact discontinuity to the north. Surface brightness profiles along and across part of the tail (green and red annular sectors) are shown in Figure~\ref{fig:tail_sur}.
We extract spectra from 4 regions along the tail (white solid box). For comparison, we extract spectra from a region adjacent to the tail (white dashed box).
We first fit these spectra to a single temperature thermal model (Figure~\ref{fig:box}). To minimize the effects of group gas projected on the tail, we employ a two temperature thermal model: {\tt vapec}$_{\rm group}$+{\tt vapec}$_{\rm tail}$. The temperature and metallicities of the {\tt vapec}$_{\rm group}$ component are fixed at the best-fits of the single temperature model of the adjacent region. The normalization (per area) of {\tt vapec}$_{\rm group}$ is allowed to vary within 20\%\footnote{The surface brightness fluctuation on a scale of $10\times10$\,kpc$^2$ in the adjacent region is 20\%.} relative to that of the adjacent region.  
The resulting best-fit parameters of {\tt vapec}$_{\rm tail}$, shown in Figure~\ref{fig:box}, reflect the intrinsic properties of the tail. 
The temperature of the tail declines from 1.3\,keV to 1\,keV, consistently cooler than the ambient group gas at 1.5\,keV. 
The Fe abundance declines from $\sim1.3$\,$Z_{\odot}$ to $\sim0.4$\,$Z_{\odot}$ along the tail and eventually reaches the Fe abundance of the ambient group gas.


\begin{figure}[h]
   \centering
       \includegraphics[width=0.5\textwidth]{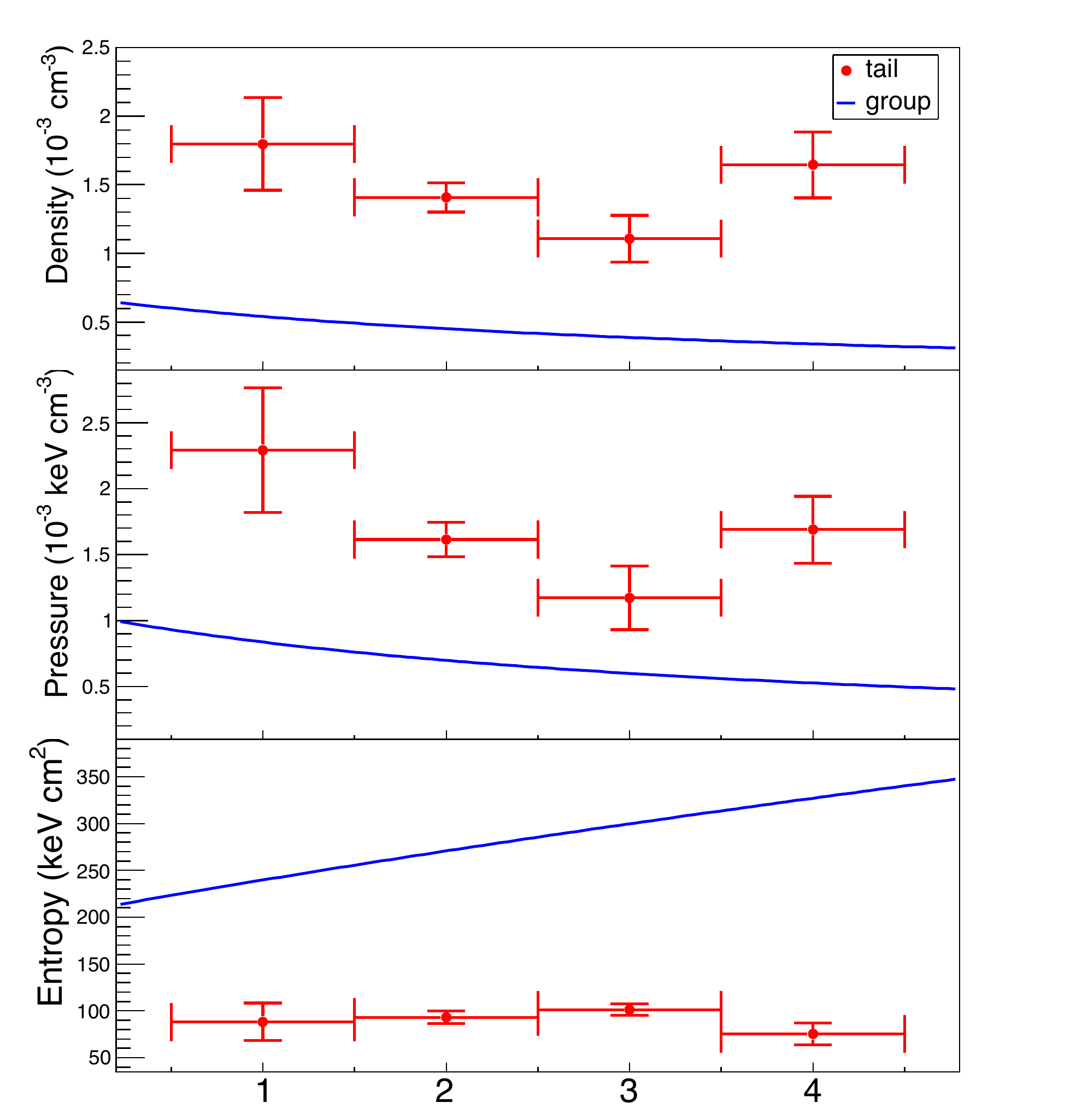}

\figcaption{\label{fig:tail3} Derived properties of the stripped tail (corresponding to the white solid box region in Figure~\ref{fig:tail}). Blue solid lines represent the group gas profiles derived from Equations (1) and (4).}
\end{figure}

The extended tail is probably shielded from the ram pressure force by the BGG such that the tail and the ambient group gas would be in pressure equilibrium. Assuming $P_{\rm tail}=P_{\rm group}$, we calculate the density and volume of the tail using the intrinsic temperature (Figure~\ref{fig:box}) and the best-fit normalization of {\tt vapec}$_{\rm tail}$. We obtain that the depth of the tail along the line of sight is 3--8 times the width of the tail. Such a pancake-shaped stripped tail is counterintuitive and unexpected from simulations.  
We then calculate the tail density assuming that the tail is cylindrical as shown in Figure~\ref{fig:tail3}. 
The resulting tail entropy ($K=kT/n^{2/3}$) and pressure are consistently lower and higher than those of the ambient group gas, respectively. 


\subsection{\bf Mass analysis}

The atmosphere of
M49 is least disturbed to the west, where the {\sl XMM-Newton} coverage also
reaches farthest from the group center, so we use this region to determine the mass
profile of the group.
The projected temperature profile centered on the BGG (corresponding to the black, cyan, and blue regions in Figure~\ref{fig:xmm}) is shown in Figure~\ref{fig:west}.
We adopt a simplified version of the analytic formula constructed by Vikhlinin et al.\ (2006) to model the deprojected temperature profile of typical cool core systems 
\begin{equation}
T_{\rm 3D}(r)=\frac{T_0}{[1+(r/r_t)^2]}\times\frac{[r/(0.075r_t)]^{1.9}+T_m/T_0}{[r/(0.075r_t)]^{1.9}+1}.
\end{equation}
The corresponding projected temperature can be obtained by projecting the gas density weighted $T_{\rm 3D}$ along the line of sight. We fit the measured projected temperature profile to the 2D formula given by Mazzotta et al.\ (2004), 
 \begin{equation}
T_{\rm 2D}=\frac{\int {{\rho}_g}^2{T_{\rm 3D}^{1/4}dz}}{\int {{\rho}_g}^2{T_{\rm 3D}^{-3/4}dz}}
 \end{equation}
 The best-fit 2D and 3D temperature profiles are plotted in Figure~\ref{fig:west}.

\begin{figure}[h]
   \centering
              \includegraphics[width=0.5\textwidth]{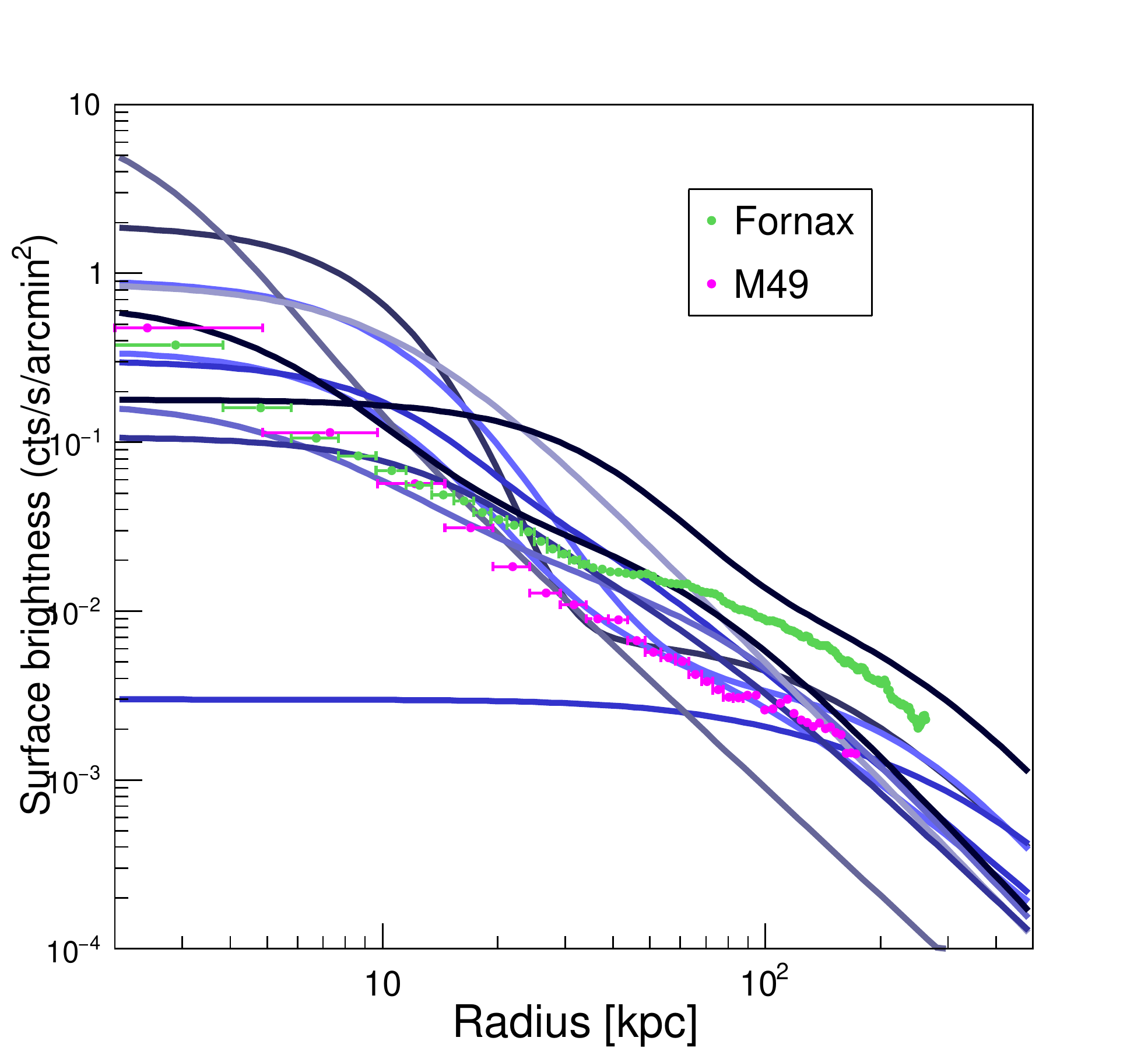}
              \figcaption{\label{fig:beta}{\sl XMM-Newton} azimuthally averaged surface brightness profile in the 0.7--2.0\,keV energy band of the M49 group, the Fornax cluster (Su et al.\ 2017d), and isolated galaxy groups (in blue) studied in Lovisari et al.\ (2015).}
              \end{figure}


We derive the 3D gas density profile following Lovisari et al.\ (2015).
We extract the surface brightness profile in the 0.7-2.0 keV energy band over the same region as the temperature profile (Figure~\ref{fig:beta}) and fit it to a double $\beta$-model which takes the form of 
\scriptsize
\begin{equation}
S(r)=S_{01}\left(1+\dfrac{r^2}{{r_{c1}}^2}\right)^{-3\beta_1+0.5}+S_{02}\left(1+\dfrac{r^2}{{r_{c2}}^2}\right)^{-3\beta_2+0.5},
\end{equation}
\normalsize
and corresponds to a density profile of 
\small
\begin{equation}
n(r)=\sqrt{{n_{01}}^2\left(1+\dfrac{r^2}{{r_{c1}}^2}\right)^{-3\beta_1}+{n_{02}}^2\left(1+\dfrac{r^2}{{r_{c2}}^2}\right)^{-3\beta_2}},
\end{equation}
\normalsize
where $n_{01}$ and $n_{02}$ are the central densities of the two components and can be related by 
\begin{equation}
\frac{n_{01}}{n_{02}}=\sqrt{\frac{S_{01}}{S_{02}}}.
\end{equation}  
To determine the central electron densities ($n_0=\sqrt{{n_{01}}^2+{n_{02}}^2}$), we fit the spectra extracted from a circular region of $R_{\rm extr}=6^{\prime}$ to 
the {\tt vapec} model; the normalization is defined as
\begin{equation}
norm=\frac{10^{-14}}{4\pi {D_A}^2(1+z)^2}\int{n_e n_H}{dV},
\end{equation}
where $D_A$ is the angular distance of the source, and $n_e$ and $n_H$ are
the electron and proton densities, for which we assume $n_e=1.21n_H$. 
We solve for $n_{01}$ and $n_{02}$ by integrating Equation (4) over
an infinite cylinder with a radius of $R_{\rm extr}$ and matching it to the best-fit normalization.

With the 3D temperature and density profiles, one can calculate the total mass within a radius $r$ assuming hydrostatic equilibrium and spherical symmetry:
\begin{equation}
M(<r)=-\frac{kTr}{G\mu m_p}\left(\frac{rd\rho}{\rho dr}+\frac{rdT}{Tdr}\right)
\end{equation}
where $k$ is the Boltzmann constant, $\mu=0.6$ is the mean atomic weight, and $G$ is the gravitational
constant. 
We randomly varied the data points of
the observed temperature and surface brightness profiles 1000 times over their statistical uncertainties. 
Each time we obtain a new set of best-fit profiles and a new total mass.
We use the standard deviation of the resulting data points as the uncertainty on mass.
The enclosed total mass profile reaches $9.2\pm0.2\times10^{12}$\,M$_{\odot}$ within 160\,kpc. 
Assuming temperature and density profiles follow Equations (2) and (4), we extrapolate 
to a virial mass of $M_{\rm vir}\approx M_{200}$ of $4.6\pm0.2\times10^{13}$\,M$_{\odot}$ out to $R_{200}=740\pm11$\,kpc. 

\section {\bf Discussion}

We have studied in detail the properties of the hot gas in the nearest infalling
galaxy group M49, which resides beyond the virial radius of Virgo.
Its temperature is below 1\,keV at the group center and rises to 1.5\,keV outside the BGG.
The group atmosphere is $>30\times$ more extended than the stellar distribution of the BGG.
Below, we discuss its dynamical state and merging history.

\subsection{\bf The northern edge: a merging cold front}

A surface brightness edge 20\,kpc north of the BGG center has been noted in previous X-ray studies (Irwin \& Sarazin 1996; Kraft et al.\ 2011).
Lacking information about the global gas distribution, previous authors
have taken this edge as the merging front between M49 and the Virgo ICM and 
determined the Mach number of the merger shock to be $\mathcal{M}=3$ based on the large discrepancy between the hot gas pressure just inside the northern edge and the expected Virgo ICM pressure at this radius. 
This Mach number corresponds to
an uncomfortably large infall velocity of $>2000$\,km\,s$^{-1}$ for a low mass cluster like Virgo, exceeding the free fall velocity from infinity. 
The likelihood\footnote{${\rm log}f=-\left(\frac{V_{\rm sub}/V_{200}}{1.55}\right)^{3.3}$, where $V_{\rm sub}$ is the infall velocity of the subcluster and $V_{\rm 200}=[10GH(z)M_{200}]^{1/3}$ is the circular velocity of the main cluster.} for this to happen is only $f\le1\times10^{-5}$ based on the velocity distribution function of dark matter substructures (Hayashi \& White 2006). 

\begin{figure}[h]
   \centering
              \includegraphics[width=0.5\textwidth]{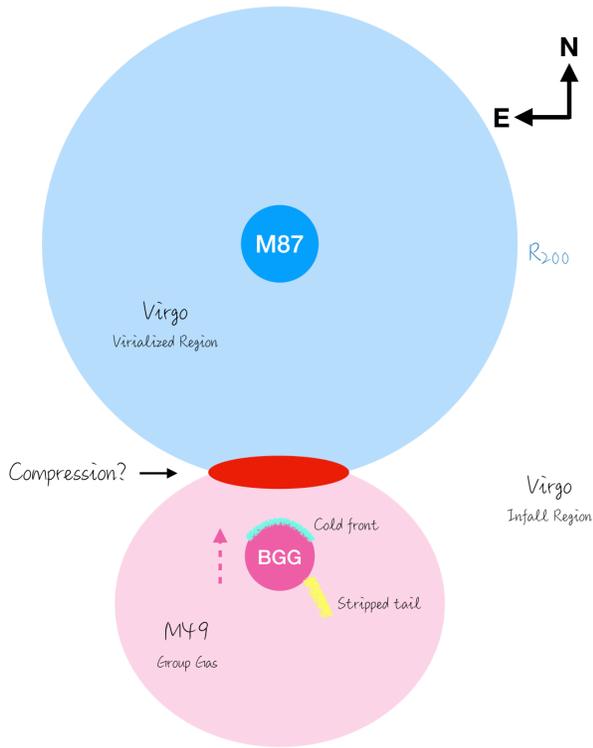}
              \figcaption{\label{fig:cartoon}A sketch demonstrates the dynamical state of the system. The encounter between the M49 group and the Virgo ICM slows down the outer layer of the M49 gas while its BGG continues moving to the north. The relative motion between the BGG and the M49 group gas gave rise to the northern cold front and the southwest stripped tail.}
              \end{figure}

Our study has shown that the northern edge is {\it not} the interface between M49 and the Virgo ICM. 
The group gas is extended to a radius of 250\,kpc to the north while this edge is at $r\sim20$\,kpc.
This interface separates the more enriched ISM of the BGG and the M49 group gas. 
Our analysis suggests that this $r\sim20$\,kpc edge has a density jump of $1.9\pm0.3$ and a temperature jump of $0.87^{+0.07}_{-0.06}$, corresponding to a pressure jump of $1.66^{+0.35}_{-0.24}$. 
We estimate the Mach number via the pressure jump (Laudau \& Lifshitz 1959):
\begin{equation}
\frac{P_0}{P_1}=\left(1+\frac{\gamma-1}{2}\mathcal{M}\right)^{\gamma/(\gamma-1)}, ~\mathcal{M}\leq 1,
\end{equation}
We take the ratio of specific heats to be $\gamma=5/3$. 
We obtain ${\mathcal{M}\approx0.67}^{+0.29}_{-0.22}$, corresponding to a velocity of $453^{+192}_{-143}$\,km\,s$^{-1}$ for a $kT\sim1.7$\,keV free stream. The inner 20\,kpc of M49 is moving subsonically northward relative to the group gas, forming a cold front. That the temperature and Fe abundance profiles vary abruptly to the north relative to those in other directions supports this scenario. As illustrated in Figure~\ref{fig:cartoon}, 
we speculate that 
this relative motion may be initiated by the encounter between the M49 group gas and the Virgo ICM, which slows down the outer layer of the M49 gas while its low entropy BGG atmosphere keeps moving to the north and makes contact with higher entropy group gas at larger radii. 


\subsection{\bf Uplift in single-phase galaxies}

{
A leap forward in understanding the fueling of super massive black holes is brought by
molecular gas detected at centers of strong cooling systems (Edge 2001). High resolution sub-millimeter observations, particularly with ALMA, reveal that such molecular clouds have a surprisingly low velocity and reside preferentially in the wake of buoyant bubbles (Russell et al.\ 2014, 2016, 2017; Vantyghem et al.\ 2016). McNamara et al. (2016) therefore propose that low-entropy gas lifted by rising bubbles becomes thermally unstable and condenses into molecular gas which could in turn fuel the AGN, dubbed as ``stimulated feedback". We observe in M49 that low entropy gas has been lifted by AGN bubbles out to 50\,kpc (\S3.2), whereas {\sl Herschel} observations of $[\rm C II]$ and $\rm H{\alpha}$ reveal a lack of cold gas in this galaxy (Werner et al.\ 2014). Another nearby massive elliptical galaxy, NGC~1399, displays similar properties (Su et al.\ 2017c; Werner et al.\ 2014).

The single-phase gas in M49 and NGC~1399 can be understood as their central cooling time substantially exceeds the free fall time ($t_{\rm cool}/t_{\rm ff}>20$), preventing precipitation throughout the cool cores (Voit et al.\ 2015). Wang et al.\ (2019) have performed a hydrodynamical simulation of the AGN feedback process specifically tailored to M49. The authors found that during each feedback cycle, its $t_{\rm cool}/t_{\rm ff}$ oscillates between 70 and 5. 
Cold gas that fuels the black hole can only form at its nucleus. 
In contrast, their simulation tailored to the multi-phase galaxy, NGC~5044, indicates a non-linear perturbation with a $t_{\rm cool}/t_{\rm ff}$ frequently reaching unity, causing its hot gas to precipitate into extended cold filaments, consistent with observations (David et al.\ 2014). In brief, for systems like M49 and NGC~1399, despite that low entropy gas has been uplifted out to large radii, 
their initial $t_{\rm cool}/t_{\rm ff}$ is sufficiently large, allowing the cooling driven AGN feedback to operate under a single-phase condition.  
}       

\subsection{\bf Ram pressure stripping and transport processes}

The BGG is moving relative to the group gas of M49, which may be responsible for the evident southwest tail originating at a distance of 50\,kpc from the group center (Figure~\ref{fig:tail}). 
We measure a negative metallicity gradient along the tail, which eventually reaches the level of the group gas.
This finding provides evidence that metals are being removed from infalling galaxies via ram pressure stripping, which could be an efficient enrichment process in the ICM (see Schindler \& Diaferio 2008 for a review). 
Stripping would occur once the ram pressure ($P_{\rm ram}\sim\rho_{\rm out}v^2$) exceeds the gravitational restoring force per unit area (Gunn \& Gott 1972). McCarthy et al.\ (2008) present a stripping condition for gaseous objects with spherical-symmetry such as early-type galaxies:
\begin{equation}
\rho_{\rm out}v^2>\frac{\pi}{2}\frac{GM_{\rm tot}\rho_{\rm in}}{R_{\rm in}}.
\end{equation}
We assume the stripping starts at the inner end of the tail, which is $50$\,kpc from the group center.
We apply an enclosed total mass of $M_{\rm tot}=2\times10^{12}$\,$M_{\odot}$ at $R_{\rm in}=50$\,kpc derived from Equation~(8) and we assume $\rho_{\rm in}\approx1.2\rho_{\rm out}$ based on the density ratio of region 1 relative to the ambient ICM (Figure~\ref{fig:tail3}-top). It would require a relative velocity of $570$\,km\,s$^{-1}$ for the stripping to take place, in broad agreement with the relative velocity of 310--645\,km\,s$^{-1}$ that we derived in \S4.1 based on the jump condition. 
The tail can be traced back to the western ghost cavity. AGN outbursts may have stirred up the ISM, lifting lower entropy gas to larger radii where the gas is less
tightly bound, facilitating the stripping process. A similar connection between AGN outbursts and stripping has been noted in M89 (Kraft et al.\ 2017; Roediger et al.\ 2015). 
 {The tail gas may originate from $r=20$\,kpc from the group center where the entropy value is the same as that of the tail. We estimate a tail gas mass of $2\times10^8$\,$M_{\odot}$. We follow Equation (11) in Su et al.\ (2017c) to calculate the minimum work required to lift such gas from $r=20$\,kpc to the inner end of the tail ($r=50$\,kpc). We obtain a minimum work of $2\times10^{56}$\,erg, comparable to the enthalpy of the ghost bubbles (Kraft et al.\ 2011).} 
Based on the estimated velocity of the BGG relative to the group gas,
it would take 100--220\,Myr to form a 70\,kpc long tail, also consistent with the age of the ghost cavities (Kraft et al.\ 2011).

The temperature of the tail declines from 1.3\,keV to 1\,keV along the tail, implying that the tail is not being heated up by the ambient group gas.
Instead, it may be cooling while expanding. The tail entropy is consistently 2--3 times smaller than the ambient group gas (Figure~\ref{fig:tail3}). 
We estimate the time scale $t_{\rm cond}$ for the tail to be erased by thermal conduction in the Spitzer regime (Spitzer 1956). The thermal energy needed is $E_1=2n_{\rm tail}\pi l^2d\times\frac{3}{2}k\Delta T$, where $l$ and $d$ are the radius and length of the tail, respectively. The heat conducted from the ICM to the tail is $E_2=qt_{\rm cond}2\pi ld$, where $q=\kappa \frac{k\Delta T}{l}$ is the heat flux and $\kappa=1.6\times10^{27}(\frac{T_{\rm ICM}}{\rm keV})^{5/2}$\,s$^{-1}$\,cm$^{-1}$ is the thermal conductivity. By equating $E_1$ and $E_2$, we obtain $t_{\rm cond}\sim5$\,Myr, which is approximately 20--40 times shorter than the estimated age of the tail. Thermal conduction may have been suppressed in the downstream region of M49.
The pressure in the tail is a factor of 2 higher than that of the ambient group gas. 
Gas in the tail may be highly inhomogeneous causing an overestimate of the mean density, which may explain the high pressure and low entropy of the tail.   

\subsection{\bf The infall velocity of M49 and the heated ICM}

The ICM temperature declines from $\sim2$\,keV to 1\,keV over a radial range of 0.5--1.5\,$R_{\rm vir}$ southward of the Virgo cluster (Figure~\ref{fig:norm}). 
{A region of elevated temperature is visible at $\approx250$\,kpc north of M49 (also see Simionescu et al.\ 2017), which could be due to a shock or adiabatic compression. 
At this stage,
if there ever was a shock propagating into the IGM of M49, it must be
well past the galaxy, allowing the flow to be responsible for the northern cold front and the downstream tail. Therefore any remaining shock at the interface between M49 and Virgo must be propagating into
Virgo. 
Assuming the temperature elevation near $R=60\arcmin$ is due to the shocked gas, we take the temperature of $\lesssim2$\,keV near $R=75\arcmin$ to be the preshock temperature (Figure~\ref{fig:norm}). 
The ICM temperature changes abruptly from $\approx2$\,keV to $3.3^{+1.2}_{-0.7}$\,keV, corresponding to a Mach number of 1.65$^{+0.60}_{-0.35}$, based on Rankine-Hugoniot shock equations (Laudau \& Lifshitz 1959):}
\begin{equation}
\frac{T_2}{T_1}=\left[\frac{2\gamma\mathcal{M}^2-(\gamma-1)}{(\gamma+1)^2}\right]\left[\gamma-1+\frac{2}{\mathcal{M}^2}\right].
\end{equation}
We estimate the expected infall velocity of M49 assuming it has experienced free fall from the turn around radius (2\,$R_{\rm vir}$) to its current position (1.25\,$R_{\rm vir}$). We assume that the dark matter halo of the cluster is truncated at $R_{\rm vir}\approx R_{200}$. The gravitational potential outside $R_{\rm vir}$ can be described by a point mass potential: 

\begin{equation}
\Phi = - \frac{4}{3} \pi G 200 \rho_c {R_{200}}^3 / r. 
\end{equation}
where $\rho_c(z)=3H(z)^2/8\pi G$ is the critical density of the Universe.
We obtain that its current infall velocity should be $\sim680$\,km\,s$^{-1}$. 
The sound speed for a 2\,keV quiescent ICM is $c_s=730$\,km\,s$^{-1}$. This leads to a Mach number of ${\mathcal{M}\lesssim1}$, falling short of producing a temperature jump of a factor of 1.65$^{+0.65}_{-0.35}$. Had M49 fallen from infinity, we would expect a Mach number of ${\mathcal{M}\sim1.3}$. 
The distance between the leading edge and the bow shock decreases as the infall velocity increases (Eq.\ (23) in Farris \& Russell 1994): 
\begin{equation}
D_{CS}=0.8R\frac{(\gamma-1)\mathcal{M}^2+2}{(\gamma+1)(\mathcal{M}^2-1),}
\end{equation}
where $R$ is the radius of a nearly spherical infalling object. 
The $\gtrsim3$\,keV region is near the boundary between the group gas and the cluster gas (Figure~\ref{fig:norm}). If the bow shock resides in this region, it would correspond to an implausibly strong shock of ${\mathcal{M}>4}$. We therefore speculate that the temperature peak may not be associated with a shock. Higher resolution observations are needed to resolve the scenario as we expect to observe a surface brightness edge near the heated ICM, in the case of a shock.

The infalling group, M49, might be able to compress some of the outer atmosphere of the Virgo cluster, causing adiabatic heating. Given a constant $P^{1-\gamma}T^{\gamma}$, the ICM pressure needs to increase by a factor of 3.5$^{+4.1}_{-1.6}$ in order to produce a temperature jump of 1.65$^{+0.60}_{-0.35}$.  
For an infall velocity of $\sim453$\,km\,s$^{-1}$ (\S4.1), we obtain 
\begin{equation}
\frac{P_f}{P_i}=\frac{P_{\rm M49}+\rho v^2}{P_{\rm Virgo}}=3.6
\end{equation}
where $P_{\rm M49}=2{n_e}kT=3.5\times10^{-4}$\,keV\,cm$^{-3}$ and $\rho=4.7\times10^{-13}$\,g\,cm$^{-3}$ are extrapolated to $r=300$\,kpc based on Equations (2) and (5) and $P_{\rm Virgo}=1.8\times10^{-4}$\,keV\,cm$^{-3}$ is the pressure of the Virgo ICM just inside its virial radius taken from Simionescu et al.\ (2017). 
It is therefore plausible for the adiabatic compression to be responsible for the temperature elevation. 

Alternatively, this $\gtrsim3$\,keV temperature could be an overestimate. After all, this temperature increase was not detected by {\sl ASCA} (Shibata et al.\ 2001). We allow the CXB component in this region free to vary and obtain a best-fit hot gas temperature of $1.5^{+0.5}_{-0.3}$\,keV, similar to the quiescent ICM; the best-fit normalization of the thermal {\tt vapec}$_{\rm ICM}$ component also falls on the extrapolation of the quiescent Virgo ICM (black solid line in Figure~\ref{fig:norm}). Meanwhile, the best-fit normalization of the CXB component is 27\% higher than what we adopted. The surface brightness 
fluctuations of unresolved point sources in the 0.5-2.0 keV band are expected to be 
$\sigma_B=3.9\times10^{-12} \Omega^{-1/2}_{0.01}$erg cm$^{-2}$ s$^{-1}$\,deg$^{-2}$,
where $\Omega_{0.01}$ is the solid angle in units of 0.01 deg$^{-2}$ 
(Miller et al.\ 2012). 
For the size of our extraction region, we expect a fluctuation of $1.9\times10^{-12}$erg cm$^{-2}$ s$^{-1}$\,deg$^{-2}$. This is 26\% of the CXB surface brightness adopted in our analysis. 
Thus the temperature enhancement might be explained by the CXB fluctuation.
 

\subsection{\bf The group atmosphere and merger history}



We compare the {\sl XMM-Newton} surface brightness profile of M49 in the 0.7--2.0 keV energy band with 
the Fornax cluster (Su et al.\ 2017d) and typical isolated galaxy groups studied in Lovisari et al.\ (2015) in Figure~\ref{fig:beta}. These systems all have average temperatures in the range of 1.0--2.0\,keV. 
The central 5--10 kpc of M49 has a similar surface brightness to other groups but it declines more quickly beyond 10--20 kpc and falls below most other isolated groups. 
This is consistent with an elevated fraction of quenched galaxies observed in the infall region compared with `field' galaxies (Boselli \& Gavazzi 2006).
If this is M49's first infall onto the outskirt of Virgo, it may have been through a severe environment outside the virial radius of Virgo. The outer layer of its group gas has been stripped off. The impact of an ICM environment on infalling objects i.e., via ram pressure stripping, can reach beyond the virial radius. 

In an alternative scenario, M49 could be a splashback group.
Gill et al.\ (2005) found that half of the galaxies in the infall region have already orbited through the main cluster. 
M49 may have just turned around at its apocenter and started to accelerate towards M87, forming a cold front and leaving behind a slingshot tail (Sheardown et al.\ 2019).
Extended X-ray mapping of the Virgo cluster reveals sloshing fronts in multiple directions and over a large span of radii, including two large scale sloshing fronts at $r=200$--300\,kpc (Simionescu et al.\ 2017). 
Roediger et al.\ (2011) investigated the sloshing scenario in Virgo through tailored simulations and identified M60, M85, and M86 as the candidates for the disturbing subclusters. However, it is difficult to reconcile their dynamical configuration and geometry. The addition of M49 would help explain the complicated thermal distribution of the Virgo ICM.
We expect future numerical simulations to resolve the orbits of M49 and the merger history of Virgo.


\section {\bf Conclusions}

M49 is a galaxy group falling onto the outskirts of the Virgo cluster.
At a distance of $\sim16$\,Mpc, this infalling group provides us the best opportunity to study the assembly of galaxy clusters beyond the virial radius. 
We present results from deep {\sl XMM-Newton} and {\sl Suzaku} mosaic observations of M49 to probe its gas properties from the group center out to radii of 150--500\,kpc. 
We see evidence that the group atmosphere slows down upon
encountering the Virgo cluster gas, causing the BGG to move forward subsonically at $v\approx450$\,km\,s$^{-1}$ relative to the group gas and creating a prominent cold front $\sim20$\,kpc north of the BGG center and a stripped tail trailing behind the central galaxy (Figure~\ref{fig:cartoon}). 

The southwest stripped tail adjacent to the BGG is 70\,kpc long (Figure~\ref{fig:tail}). The tail temperature declines outward along the tail, suggesting that the tail is not being heated up by the ambient group gas but is cooling via expansion. The metallicity of the tail declines from $\sim1.3$\,$Z_{\odot}$ to $\sim0.4$\,$Z_{\odot}$ with distance from the galaxy center and eventually reaches the level of the group gas, 
supporting the case that ram pressure stripping could be effective at enriching the intracluster medium. AGN outbursts in the BGG may have facilitated the stripping process. 
Inhomogeneities cause the mean gas density to be overestimated, which
might account for the measured tail entropy and pressure being lower
and higher, respectively, to those of the ambient gas.

Although M49 has an extended group halo, its atmosphere appears truncated when compared with isolated galaxy groups (Figure~\ref{fig:beta}). 
If this is M49's first infall, our findings imply that 
the Virgo ICM extends well beyond the virial radius and provides a severe environment for infalling objects. 
Alternatively, M49 may be a splashback group, so that its atmosphere was
truncated during its passage within the virial radius of Virgo. Its merger history may help explain the complicated sloshing structure observed in the Virgo ICM. 



\section{\bf Acknowledgments}
We thank the anonymous referee for helpful comments. We thank Yuan Li for helpful discussions during her visit at the University of Kentucky.  
Y.S. acknowledges support from {\sl Chandra} Awards GO6-17125X issued by the
{\sl Chandra} X-ray Observatory Center which is operated by the Smithsonian Astrophysical Observatory. F.M. acknowledges support from the Lend\"ulet LP2016-11 grant awarded by the Hungarian Academy of Sciences.


\begin{references}
\reference{} Arnaud K. A., 1996, in Jacoby G. H., Barnes J., eds, Astronomical Society of the Pacific Conference Series Vol. 101, Astronomical Data Analysis Software and Systems V. p. 17
\reference{} Anders, E. \& Grevesse, N. 1989, GeCoA, 53, 197
\reference{} Asplund, M., Grevesse, N., Jacques S.\ A. 2009, ARA\&A, 47, 481
\reference{} Berrier, J.\ C., Stewart, K.\ R., Bullock, J.\ S. et al.\ 2009, ApJ, 690, 1292
\reference{} Bianconi, M., Smith, G.\ P., Haines, C.\ P. et al.\ 2018, MNRAS, 473, 79
\reference{} Biller, B.\ A., Jones, C., Forman, W.\ R. et al.\ 2004, ApJ, 613, 238
\reference{} Blakeslee, J. P., Jordan, A., Mei, S., et al.\ 2009, ApJ, 694, 556
\reference{} B\"ohringer, H.; Briel, U. G.; Schwarz, R. A. et al. 1994, Natur, 368, 828
\reference{} Boselli, A. \& Gavazzi, G. 2006, PASP, 118, 517
\reference{} Boylan-Kolchin M., Springel V., White S.\ D.\ M., et al.\ 2009, MNRAS, 398, 1150
\reference{} Butcher, H. \& Oemler, A., Jr.\ 1978, ApJ, 226, 559
\reference{} Cappellari, M.\ \& Copin, Y. 2003, MNRAS, 342, 345 
\reference{} Cappellari, M., Emsellem, E., Krajnovic, D. et al.\ 2011, MNRAS, 416, 1680
\reference{} David, L. P., Lim, J., Forman, W., et al.\ 2014, ApJ, 792, 94
\reference{} De Luca, A. \& Molendi, S. 2004, A\&A, 419, 837
\reference{} De Grandi, S., Eckert, D., Molendi, S. et al.\ 2016, A\&A, 592, 154
\reference{} Dickey, J.\ M.\ \& Lockman, F.\ J.\ 1990, ARA\&A, 28, 215
\reference{} Diehl, S. \& Statler, T.\ S. 2006, MNRAS, 368, 497
\reference{} Dressler, A. 1980, ApJ, 326, 351
\reference{} Eckert, D., Molendi, S., Owers, M., et al.\ 2014, A\&A, 570, 119
\reference{} Eckert, D., Gaspari, M., Owers, M.\ S. et al.\ 2017, A\&A, 605, 25
\reference{} Edge, A.\ C.\ 2001, MNRAS, 328, 762
\reference{} Ettori, S., Gastaldello, F., Leccardi, A., et al.\ 2010, A\&A, 524, A68 
\reference{} Farris, M. H., \& Russell, C. T. 1994, J. Geophys. Res., 99, 17
\reference{} Forman, W., Jones, C., David, L. et al.1993 ApJ 418L 55
\reference{} Fujita Y. 2004, PASJ, 56, 29
\reference{} Gendron-Marsolais, M., Kraft, R.\ P., Bogdan, A. et al.\ 2017, ApJ, 848, 26
\reference{} Gill, S.\ P.\ D., Knebe, A., Gibson, B.\ K. 2005, MNRAS, 356, 1327 
\reference{} Gunn, J.\ E. \& Gott, J.\ R. III 1972, ApJ, 176, 1 
\reference{} Hayashi, E. \& White, S.\ D.\ M. 2006, MNRAS, 370, 38
\reference{} Haines, C.\ P., Finoguenov, A., Smith, G.\ P. et al.\ 2018, MNRAS, 447, 4931
\reference{} Ichinohe, Y., Werner, N., Simionescu, A. et al.\ 2015, MNRAS, 448, 2971
\reference{} Irwin, J.\ A. \& Sarazin, C.\ L. 1996, ApJ, 471, 683
\reference{} Irwin, J.\ A., Athey, A.\ E., Bregman, J.\ N. 2003, ApJ, 587, 356
\reference{} Kraft, R.\ P., Forman, W.\ R., Jones, C. et al.\ 2011, ApJ, 727, 41
\reference{} Kraft, R.\ P., Roediger, E., Machacek, M. et al.\ 2017, ApJ, 848, 27
\reference{} Lakhchaura, K., Mernier, F., Werner, N.\ 2019, A\&A, 623, 17
\reference{} Landau L.\ D., Lifshitz E.\ M., 1959, Fluid mechanics. Oxford, Pergamon Press
\reference{} Lodders, K. 2003, ApJ, 591, 1220 
\reference{} Lovisari, L., Reiprich, T.\ H., Schellenberger, G. 2015, A\&A, 573, 118 
\reference{} McNamara, B.\ R., Russell, H.\ R., Nulsen, P.\ E.\ J., et al.\ 2016, ApJ, 830, 79
\reference{} Mazzotta, P., Rasia, E., Moscardini, L., et al.\ 2004, MNRAS, 354, 10
\reference{} McCammon, D., Almy, R., Apodaca, E., et al.\ 2002, ApJ, 576, 188
\reference{} McCarthy, I.\ G., Frenk, C.\ S., Font, A.\ S. et al.\ 2008, MNRAS, 383, 593
\reference{} Mei, S., Blakeslee, J.\ P., Cote, P. et al.\ 2007, ApJ, 655, 144
\reference{} Mernier, F., de Plaa, J., Pinto, C. et al.\ 2016, A\&A, 592, 157
\reference{} Mernier, F., de Plaa, J., Kaastra, J.\ S. et al.\ 2017, A\&A, 603, 80
\reference{} Miller, E.\ D., Bautz, M., George, J. et al.\ 2012, AIPC, 1427, 13
\reference{} Molendi, S., De Luca, A. \& Leccardi, A.\ 2004, A\&A 419, 837
\reference{} More, S., Diemer, B., Kravtsov, A.\ V. 2015, ApJ, 810, 36
\reference{} Neumann, D.\ M., Lumb, D.\ H., Pratt, G.\ W., et al.\ 2003,
A\&A, 400, 811
\reference{} Oemler, A., Jr. 1974, ApJ, 194, 1 
\reference{} Panagoulia, E.\ K., Fabian, A.\ C., Sanders, J.\ S.\ 2013, MNRAS, 433, 3290
\reference{} Panagoulia, E.\ K., Sanders, J.\ S., Fabian, A.\ C.\ 2015, MNRAS, 447, 417
\reference{} Roediger, E., Br$\ddot{\rm u}$ggen, M., Simionescu, A. et al.\ 2011, MNRAS, 413, 2057
\reference{} Roediger, E., Kraft, R.\ P., Nulsen, P.\ E.\ J. et al.\ 2015, ApJ, 806, 104
\reference{} Russell, H.\ R., McNamara, B.\ R., Edge, A.\ C., et al.\ 2014, ApJ, 784, 78
\reference{} Russell, H.\ R., McNamara, B.\ R., Fabian, A.\ C., et al.\ 2016, MNRAS,
458, 3134
\reference{} Russell, H.\ R., McDonald, M., McNamara, B.\ R., et al.\ 2017, ApJ, 836, 130
\reference{} Sanders, J.\ S., Fabian, A.\ C., Taylor, G.\ B. et al.\ 2016, MNRAS, 457, 82
\reference{} Schindler, S \& Diaferio, A. 2008, SSRv, 134, 363
\reference{} Sheardown, A., Fish, T.\ M., Roediger, E. et al.\ 2019, ApJ, 874, 2
\reference{} Shibata, R., Matsushita, K., Yamasaki, N.\ Y. et al.\ 2001, ApJ, 549, 228
\reference{} Simionescu, A., Werner N., Urban, O., et al.\ 2015, ApJ, 811, 25L
\reference{} Simionescu, A., Werner, N., Mantz, A. et al.\ 2017, ApJ, 469, 1476
\reference{} Spitzer 1956, Physics of Fully Ionized Gases (New York: Interscience Publishers)
\reference{} Snowden S. L.,  Kuntz K. D., Cookbook for Analysis Procedures for XMM–Newton EPIC MOS Observations of Extended Objects and the Diffuse Background , 2011 ftp://xmm.esac.esa.int/pub/xmm-esas/xmm-esas.pdf
\reference{} Su, Y., White, R.\ E., III, Miller, E.\ D. ApJ, 2013, 775, 89 
\reference{} Su, Y., Gu, L., White, R. et al.\ 2014, ApJ, 786, 152
\reference{} Su, Y., Buote, D., \& Gastaldello, F.\ et al.\ ApJ 2015, 805, 104
\reference{} Su, Y., Kraft, R.\ P., Nulsen, P.\.E.\ J. et al.\ 2017a, ApJ, 835, 19
\reference{} Su, Y., Kraft, R.\ P., Roediger, E. et al.\ 2017b, ApJ, 834, 74
\reference{} Su, Y., Nulsen, P.\ E.\ J.,  Kraft, R.\ P., et al.\ 2017c, ApJ, 847, 94
\reference{} Su, Y., Nulsen, P.\ E.\ J.,  Kraft, R.\ P., et al.\ 2017d, ApJ, 851, 69
\reference{} Temi, P., Brighenti, F., Mathews, W.\ G. 2007, ApJ, 660, 1215
\reference{} Urban, O., Werner, N., Simionescu, A. et al.\ 2011, MNRAS, 414, 2101
\reference{} Vantyghem, A.\ N., McNamara, B.\ R., Russell, H.\ R., et al.\ 2016, ApJ, 832, 148
\reference{} Vikhlinin, A., Markevitch, M., Murray, S. S. 2001, ApJ, 551, 160
\reference{} Vikhlinin, A., Kravtsov, A., Forman, W. et al.\ 2006, ApJ, 640, 691
\reference{} Voit, G.\ M., Donahue, M., O’Shea, B.\ W., et al.\ 2015, ApJL, 803, L21
\reference{} Wang, C., Li, Y., Ruszkowski, M. 2019, MNRAS, 482, 3576
\reference{} Werner, N., Oonk, J.\ B.\ R., Sun, M., et al.\ 2014, MNRAS, 439, 2291
\reference{} Zabludoff A.\ I., Zaritsky D., Lin H. et al.\ 1996, ApJ, 466,
104
\end{references}
\end{document}